%% file: _main.tex
\input{_constants}
\arxiv

\pdfoutput=1
\documentclass[10pt,twocolumn,letterpaper]{article}
\input{cvpr_header}

\makeatletter
\renewcommand{\l@section}{\@dottedtocline{1}{1.5em}{3em}}
\renewcommand{\l@subsection}{\@dottedtocline{2}{0.0em}{3.5em}}
\makeatother

\begin{document}
\title{\paperTitle}

\author{Marius Schmidt-Mengin$^{1}$ \quad Alexis Benichoux$^{1}$ \quad Shibeshih Belachew$^{1}$ \\   Nikos Komodakis$^{1,3,4,5}$ \quad Nikos Paragios$^{1,2}$ \\
{\normalsize $^1$TheraPanacea, France} \quad
{\normalsize $^2$CentraleSupélec, University of Paris-Saclay, France}  \quad \\
{\normalsize $^3$University of Crete, Greece}  \quad 
{\normalsize $^4$IACM, FORTH, Greece}  \quad 
{\normalsize $^5$Archimedes, Athena RC, Greece}  \quad \\
}

\maketitle

\input{00_abstract}
\input{01_intro}
\input{02_related}

\input{03_method}
\input{04_experiments}

\input{10_conclusion}

{\small
\bibliographystyle{ieeenat_fullname}
\bibliography{11_references}
}

\ifreview \clearpage
\appendix
\input{12_appendix}

\fi

\ifcamera \clearpage
\appendix \input{12_appendix} 
\fi

\ifarxiv \clearpage
\appendix \input{12_appendix} 
\fi
\end{document}

%% file: _constants.tex
\def\paperID{6359}
\def\confName{CVPR}
\def\confYear{2024}

\def\paperTitle{ToNNO: Tomographic Reconstruction of a Neural Network's Output \\ for Weakly Supervised Segmentation of 3D Medical Images}

\newif\ifreview 
\newif\ifarxiv \newcommand{\arxiv}{\arxivtrue}
\newif\ifcamera 
\newif\ifrebuttal 

%% file: cvpr_header.tex
\ifreview \usepackage[review]{cvpr} \fi
\ifarxiv \usepackage[pagenumbers]{cvpr} \fi
\ifrebuttal \usepackage[rebuttal]{cvpr} \fi
\ifcamera \usepackage{cvpr} \fi

\input{_macros}  

\usepackage{xr-hyper}

\makeatletter
\newcommand*{\addFileDependency}[1]{
  \typeout{(#1)}
  \@addtofilelist{#1}
  \IfFileExists{#1}{}{\typeout{No file #1.}}
}

\makeatother

\definecolor{cvprblue}{rgb}{0.21,0.49,0.74}
\usepackage[pagebackref,breaklinks,colorlinks,citecolor=cvprblue]{hyperref}
\usepackage[capitalize]{cleveref}
\crefname{section}{Sec.}{Secs.}
\crefname{table}{Table}{Tables}
\crefname{figure}{Fig.}{Figs.}

%% file: _macros.tex
\usepackage{subcaption}
\usepackage{graphicx}	
\usepackage{amsmath}	
\usepackage{amssymb}	
\usepackage{booktabs}
\usepackage{times}
\usepackage{microtype}
\usepackage{epsfig}
\usepackage[table,xcdraw,dvipsnames]{xcolor}
\usepackage{caption}
\usepackage{float}
\usepackage{placeins}
\usepackage{color, colortbl}
\usepackage{stfloats}
\usepackage{enumitem}
\usepackage{tabularx}
\usepackage{xstring}
\usepackage{multirow}
\usepackage{xspace}
\usepackage{url}
\usepackage[hang,flushmargin]{footmisc}
\usepackage{mathtools}
\usepackage{makecell}
\usepackage{anyfontsize}
\usepackage[none]{hyphenat}
\usepackage{dsfont}
\usepackage{adjustbox}
\usepackage[dvipsnames]{xcolor}
\usepackage{titletoc}
\usepackage[accsupp]{axessibility}

\newcommand\DoToC{%
  \startcontents
  \printcontents{}{1}{\textbf{Contents}\vskip3pt\hrule\vskip5pt}
  \vskip3pt\hrule\vskip5pt
}

\DeclareCaptionFont{figurefont}{\fontsize{6.5}{6.5}\selectfont}
\ifcamera \usepackage[accsupp]{axessibility} \fi

% ...

\ifarxiv  \fi

\newcommand{\R}[1]{{%
    \textbf{%
        \ifstrequal{#1}{1}{\textcolor{red}{R#1}}{%
        \ifstrequal{#1}{2}{\textcolor{blue}{R#1}}{%
        \ifstrequal{#1}{3}{\textcolor{magenta}{R#1}}{%
        \ifstrequal{#1}{4}{\textcolor{teal}{R#1}}{%
                           \textcolor{cyan}{R#1}%
        }}}}%
    }%
}}

\newcommand{\uv}[1]{\widehat{\mathbf{#1}}}

%% file: 00_abstract.tex
\begin{abstract}
Annotating lots of 3D medical images for training segmentation models is time-consuming.
The goal of weakly supervised semantic segmentation is to train segmentation models without using any ground truth segmentation masks. Our work addresses the case where only image-level categorical labels, indicating the presence or absence of a particular region of interest (such as tumours or lesions), are available.
Most existing methods rely on class activation mapping (CAM). We propose a novel approach, \textbf{ToNNO}, which is based on the \textbf{To}mographic reconstruction of a \textbf{N}eural \textbf{N}etwork's \textbf{O}utput. Our technique extracts stacks of slices with different angles from the input 3D volume, feeds these slices to a 2D encoder, and applies the inverse Radon transform in order to reconstruct a 3D heatmap of the encoder’s predictions. This generic method allows to perform dense prediction tasks on 3D volumes using any 2D image encoder. We apply it to weakly supervised medical image segmentation by training the 2D encoder to output high values for slices containing the regions of interest. We test it on four large scale medical image datasets and outperform 2D CAM methods. We then extend ToNNO by combining tomographic reconstruction with CAM methods, proposing Averaged CAM and Tomographic CAM, which obtain even better results.
\end{abstract}

%% file: 01_intro.tex
\section{Introduction}
\label{sec:intro}

\begin{figure}[ht] 
   \begin{subfigure}{0.32\linewidth}
       \includegraphics[width=\linewidth, trim={0 0.75cm 0 1.3cm}, clip]{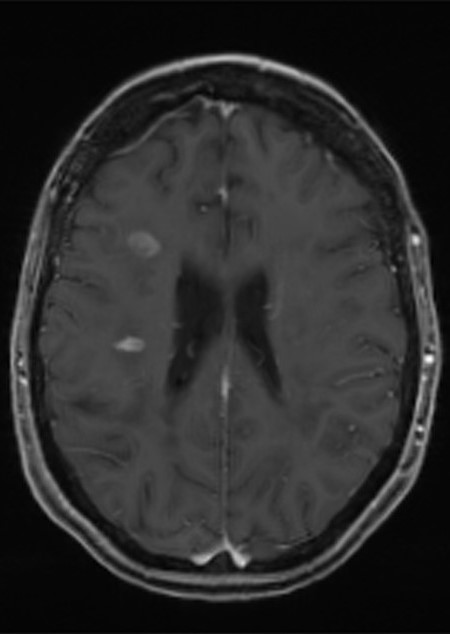}
       \caption{Input}
   \end{subfigure}
\hfill 
   \begin{subfigure}{0.32\linewidth}
       \includegraphics[width=\linewidth, trim={0 0.75cm 0 1.3cm}, clip]{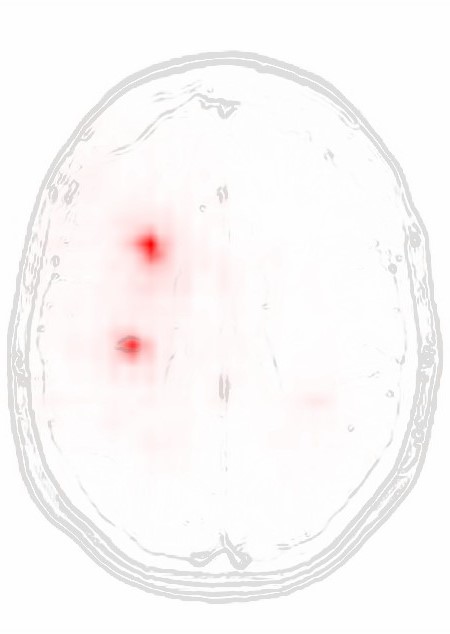}
       \caption{LayerCAM \cite{jiang2021layercam}}
   \end{subfigure}
\hfill 
   \begin{subfigure}{0.32\linewidth}
       \includegraphics[width=\linewidth, trim={0 0.75cm 0 1.3cm}, clip]{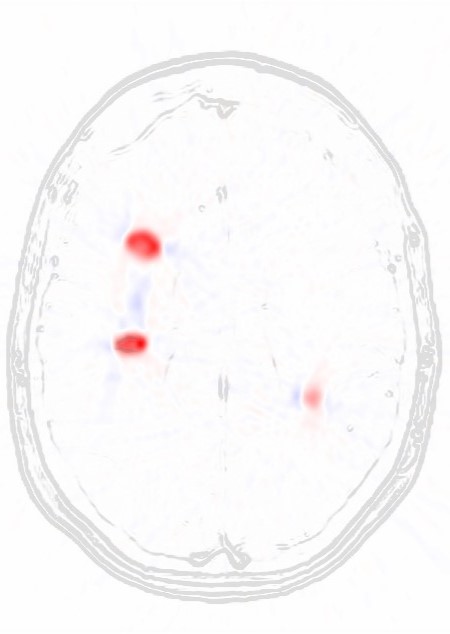}
       \caption{ToNNO}
   \end{subfigure}

   \caption{Our method, which allows to obtain high resolution segmentations  without using any ground truth segmentation masks, is completely orthogonal to class activation mapping methods such as GradCAM \cite{selvaraju2017grad} or LayerCAM \cite{jiang2021layercam}.}
   \label{fig:first_image}
\end{figure}

The advent of 3D medical imaging, such as computed tomography (CT), magnetic resonance imaging (MRI) or positron emission tomography (PET) has revolutionized clinical practice by enabling precise visualization and quantitative analysis of the internal structures of the body. Many scenarios involve segmenting particular regions of interest. For example, in radiotherapy workflows, it is necessary to segment organs and tumours in order to compute a treatment plan that maximizes the radiation dose received by the tumours while sparing the surrounding healthy tissues. In multiple sclerosis, segmentation of brain lesions allows to estimate lesion load for monitoring patients \cite{icobrain}. Furthermore, automatic segmentation methods may be useful as a verification step in order to detect tumours or lesions that were missed by human experts.

However, the volumetric nature of 3D medical images (also referred to as volumes) makes their manual segmentation slow and painstaking. Deep learning has become an important tool to automate this task. Deep learning techniques, though, typically still require large amounts of manually segmented training data, which can be prohibitive. Furthermore, manual segmentation can be subjective and vary a lot between different experts. Weakly supervised semantic segmentation methods aim to circumvent this issue by training semantic segmentation models without using any ground truth segmentation masks, relying instead on bounding boxes, scribbles, point-wise annotations, or image-level labels. Our work focuses on the latter.

To this end, we propose \textbf{ToNNO} (\textbf{To}mographic reconstruction of a \textbf{N}eural \textbf{N}etwork's \textbf{O}utput), a generic method that produces a 3D heatmap which represents the predictions of a 2D encoder (such as a ResNet \cite{resnet}) for different slices of a given input volume. While many methods adopt an encoder-decoder structure, our method allows to perform dense prediction tasks on 3D images using only a 2D encoder. Furthermore, it enables to tap into the large amount of available 2D neural network implementations and pretrained weights. Our method is inspired by computed tomography. We extract 2D slices from a 3D image at many different angles, feed each of them to the 2D encoder, and use a tomographic reconstruction technique, namely the inverse Radon transform, in order obtain a volumetric heatmap that represents the predicted slice logits in 3D. 

We then apply this method to weakly supervised medical image semantic segmentation.
Using only binary image-level labels that indicate whether or not the 3D images contain a given region of interest (e.g., tumour, lesion, etc.), we first train a 2D classifier to predict whether or not individual slices of these images contain the region of interest.
Then, we apply ToNNO and show that it allows us to obtain a high resolution segmentation of the regions of interest, as shown in \Cref{fig:first_image}.

ToNNO is orthogonal to class activation mapping (CAM) \cite{cam}, which is currently the most common family of methods for weakly supervised medical image segmentation.
As a baseline, we apply GradCAM \cite{selvaraju2017grad} and LayerCAM \cite{jiang2021layercam} to the 2D classifier that we trained for ToNNO. We find that in most cases, ToNNO outperforms these two methods.
Using the ideas behind ToNNO, we also propose to average the class activation maps produced by GradCAM \cite{selvaraju2017grad} and LayerCAM \cite{jiang2021layercam} across many different angles, boosting the results of these methods by large amounts. Furthermore, we find that incorporating the filtering step---a key ingredient of the tomographic reconstruction technique that we use---into the averaging process allows to correct the inherent blurriness of class activation maps in order to obtain sharp averaged CAM heatmaps even for the deepest layers.

Thus, our contribution is threefold. First, we propose a framework to train a 2D classifier on slices extracted at any angle from the 3D volumes, using volume-level labels. Second, we propose ToNNO, a novel method to reconstruct 3D segmentations using this trained 2D classifier. Third, we incorporate CAM methods into our reconstruction framework, proposing Averaged CAM and Tomographic CAM.

%% file: 02_related.tex
\section{Related Work}
\label{sec:related}

\paragraph{Weakly supervised semantic segmentation.} This paradigm aims to segment 2D or 3D images without using ground truth segmentation masks. Multiple forms of supervision can be used instead: orthogonal segmentation masks \cite{ortho_annot}, bounding boxes \cite{grabcut, weakly_and_semi_2015, boxsup, simply_does_it, 3d_guided, boxinst, box2mask}, scribbles \cite{inter_graph_cut, scribble_rand_walk, scribblesup, cyclemix, trimix, scribble_polyp}, point-wise annotations \cite{ms_pointwise, inter_graph_cut, weakly_histo_point, whats_the_point}, or, in its weakest and most challenging form, image-level labels \cite{from_image_level_cnns, seed_expand_constrain, weakly_affinity_3d, image_level_pix_affinity}. When using image-level labels, most methods are based on deep learning and class activation mapping (CAM) \cite{cam, selvaraju2017grad, gradcam_pp, eigen_cam, score_cam, jiang2021layercam}. They make use of a classifier that was trained to classify between the different image-level labels and exploit activations and/or gradients of intermediate layers in order to obtain spatial information about the regions that cause the classifier to make its decision.
For medical weakly supervised segmentation, CAM methods are most often applied to 2D slices \cite{weakly_mri_soft_bounds, weakly_pneumo}, sometimes deriving slice-level labels from 3D segmentation masks \cite{ame_cam, causal_cam, weakly_autopet}. In this work, we apply GradCAM \cite{selvaraju2017grad}, and LayerCAM \cite{jiang2021layercam} which was shown to outperform GradCAM by large margins.
One drawback of CAM methods is that the heatmaps that they generate usually have a resolution that is lower than the input.

\paragraph{3D medical image segmentation.} It can be achieved with 3D encoder-decoder style neural networks \cite{vnet}, which maintain the 3D nature of the input data throughout the depth of the model. But 3D models can be hard to fit into the memory of GPUs for high resolution input data \cite{fastsurfer}. Thus, many of these methods rely on three 2D models that produce segmentations for each slice of the input volume in each viewing direction (axial, sagittal and coronal), which are then assembled back into a 3D segmentation 
\cite{fastsurfer, quicknat, vesselnet, fusionnet, airway, cascaded, pulmonary}. In contrast, our method trains a single 2D encoder for slices which can have any 3D orientation.



\paragraph{Tomography.} Tomographic reconstruction allows to recover volumetric information from slice-wise information. It is formalized by the Radon transform \cite{radon_uber_1917, Radon_book} and most importantly, its inverse. Applications of tomographic reconstruction techniques include computed tomography (CT) imaging and electron tomography (ET). However, CT imaging is based on the 2D-1D Radon transform, meaning that 2D cross-sections of the patient are reconstructed using 1D integrals of the absorption of X-rays, and ET is based on either the 2D-1D or the 3D-1D Radon transform. These cases of the Radon transform have been extensively studied \cite{ct_book, ct_dl, ct_dl_book, et_book}. Our work, however, uses the rather uncommon 3D-2D Radon transform, where a 3D volume is reconstructed using 2D slice integrals. To the best of our knowledge, no previous work has combined deep learning with the inverse Radon transform for dense prediction tasks on 3D images.

%% file: 03_method.tex
\section{Method}
\label{sec:method}

\begin{figure}[t]
  \centering
  \begin{subfigure}[b]{1\linewidth}
       \includegraphics[width=1\linewidth]{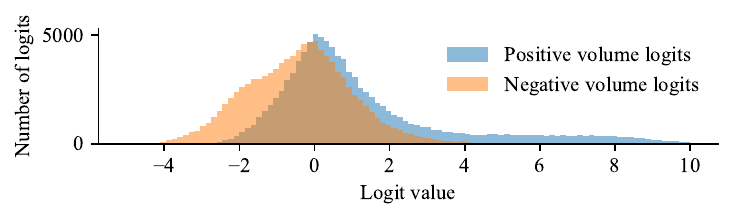}
        \vspace{-15pt}
        \caption{Histograms of logits for 100,000 slices of the MosMed dataset (validation set) extracted from multiple positive and negative volumes.}
        \label{fig:logit_hist_top}
   \end{subfigure}

   \begin{subfigure}[b]{\linewidth}
       \includegraphics[width=1\linewidth]{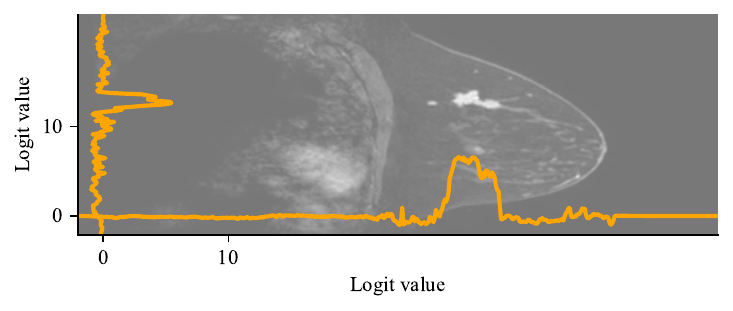}
       \vspace{-15pt}
        \caption{Logits produced by the trained classifier for each slice of the volume shown in the background (Duke dataset) along the vertical and horizontal axis allow to pinpoint the location of the tumour.
        }
        \label{fig:logit_hist_bottom}
   \end{subfigure}
   \caption{Visualisation of logits produced by the 2D classifier  $g_\theta$}

\end{figure}

\subsection{Intuition}

We assume that we have a dataset of volumes with associated binary labels $\{(V_1, y_1), ..., (V_N, y_n)\}$, where $\forall i, {V_i \in \mathbb{R}^{d_i\times h_i \times w_i}}$ and $y_i \in \{0, 1\}$. 
Our goal is to obtain a segmentation of the regions that distinguish the positive class from the negative class. For example, if the positive class consists of images of cancer patients and the negative class consists of images of healthy patients, our goal is to segment the tumours. Our method starts by training a 2D classification neural network $g_\theta$, with parameters $\theta$, to distinguish between slices of positive volumes and slices of negative volumes.
In order to do that, we sample random slices (which can have any 3D orientation) from the training volumes and associate to each slice the ground truth label of the volume from which it came. Note that this effectively introduces label noise, as in positive volumes, all slices get assigned a positive label even if they don't contain any regions of interest. But we have no way of deriving more precise labels, as per the problem definition, only volume-level labels are available. We then optimize $\theta$ to minimize the binary cross-entropy loss of each slice with respect to its assigned label.

After training, the distributions of the logits produced by the classifier for slices coming from the positive and negative classes are different, as shown in \Cref{fig:logit_hist_top}.
When feeding each slice of a given volume to the classifier, we obtain a logit profile along the slice axis that can allow us to pinpoint the location of the regions of interest along this axis (see \Cref{fig:logit_hist_bottom}). If we repeat this process for different slice axes, we can narrow down the location in three dimensions. This process is formalized by the theory of tomographic reconstruction, whose principal mathematical tool is the Radon transform.

Our weakly supervised segmentation method thus consists in two steps. First, we train the classifier with stacks of slices that can have any 3D orientation, which can be seen as applying random 3D rotations to the volumes during training. Then, we reconstruct a heatmap by applying the inverse Radon transform to logits obtained for stacks of slices across many different orientations.

\begin{figure*}[ht]
    \centering
    \includegraphics[width=1\textwidth]{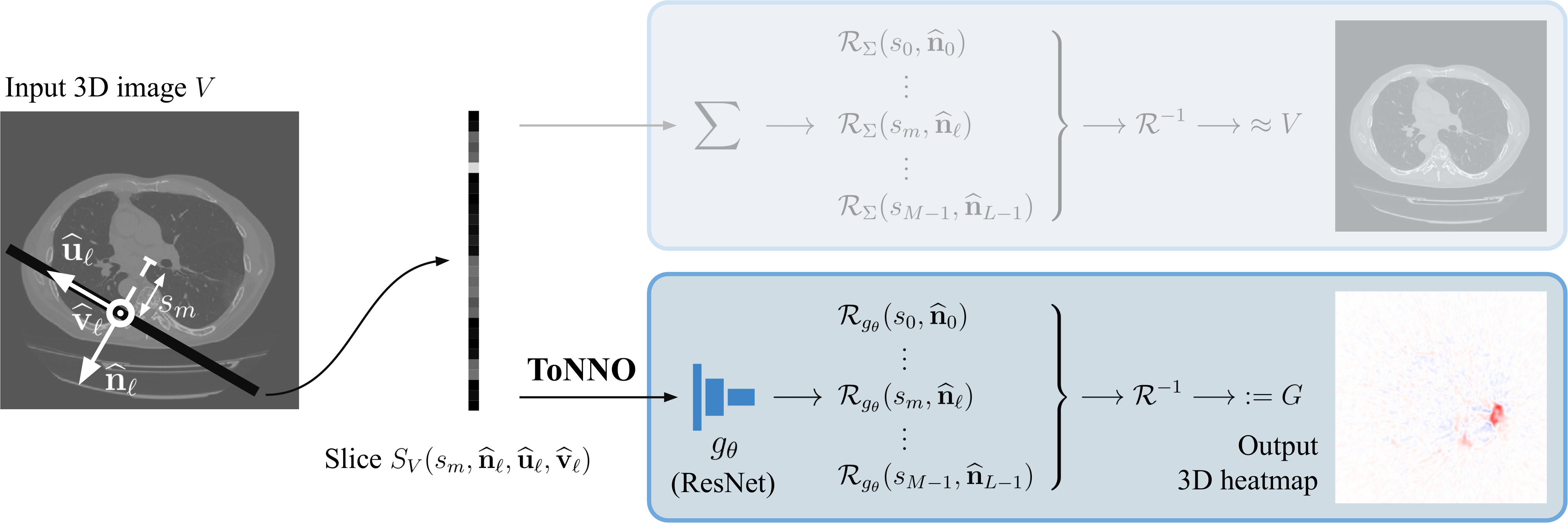}
    \caption{\textbf{Overview of ToNNO.} 3D volumes are represented as 2D images and 2D slices as 1D rows of pixels. First, slices are extracted from the input volume (left). By summing the pixels of each slice (top), we can approximate the Radon transform of the input volume $V$, and applying the inverse Radon transform $\mathcal{R}^{-1}$ allows to reconstruct $V$. The idea of this work is to replace the sum by a trained neural network before applying the inverse Radon transform (bottom).}
    \label{fig:main_diagram}
\end{figure*}

\subsection{Radon transform and its inverse}

Let $f: \mathbb R^3 \rightarrow \mathbb R$ be a 3D function. For the sake of simplicity, we assume it to be infinitely differentiable with compact support. Let $\mathbb S^2$ be the unit sphere of $\mathbb R^3$. For a unit vector $\uv n \in \mathbb S^2$ and $s \in  \mathbb R$, let $\xi(s, \uv n) = \left\{\mathbf x \in \mathbb R^3, \mathbf x \cdot \uv n = s\right\}$ be the plane orthogonal to $\uv n$ containing the point $s\uv n$. We define the 3D-2D Radon transform of $f$ as the function $\mathcal R f: \mathbb R \times \mathbb S^2 \rightarrow \mathbb R$ by

\begin{equation}
    \label{eq:radon_transform}
    \mathcal R f(s, \uv n) \coloneqq \int_{\mathbf x \in \xi(s, \uv n)} f(\mathbf x) \;\mathrm d\mathbf x
\end{equation}
It can be shown \cite[Theorem 3.6]{Radon_book} that $f$ can be recovered from $\mathcal R f$ using the filtered backprojection formula:
\begin{equation}
 f(\mathbf x) = c \int_{\uv n \in \mathbb S^2} \frac{\partial^2 \mathcal R f}{\partial s^2}(\mathbf x \cdot \uv n, \uv n) \; \mathrm d \uv n
\end{equation}
where $c$ is a negative constant.
For a function ${\varphi: \mathbb R \times \mathbb S^2 \rightarrow \mathbb R}$, we thus define ${\mathcal{R}^{-1} \varphi: \mathbb R^3 \rightarrow \mathbb R}$ by
\begin{equation}
\label{eq:inv_radon_transform}
    \mathcal R^{-1} \varphi(\mathbf x) \coloneqq c \int_{\uv n \in \mathbb S^2} \frac{\partial^2 \varphi}{\partial s^2}(\mathbf x \cdot \uv n, \uv n) \; \mathrm d \uv n
\end{equation}
It holds that for all $\uv n$, $\mathbf x \in \xi(\mathbf x \cdot \uv n, \uv n)$. Thus, for a given $\mathbf x$, this integral integrates the function $\partial^2\varphi/\partial s^2$ over all planes containing 
the point $\mathbf x$ and is called the backprojection operator.

\subsection{Implementation}
Now, let
$V: [d] \times [h] \times [w] \rightarrow \mathbb R$
(where for an integer $n$, we define $[n] = \{0, 1, ..., n-1\}$) be a discrete 3D volume, such as a medical image. The Radon transform cannot be directly applied to $V$, as $V$ is not defined on $\mathbb R^3$. Let $\widetilde V: \mathbb R^3 \rightarrow \mathbb R$ be an interpolation of $V$, such that $\widetilde V(-1, -1, -1) = V(0, 0, 0)$ and $\widetilde V(1, 1, 1) = V(d-1, h-1, w-1)$, and $\widetilde V(\mathbf x) = 0$ if $\mathbf x \notin [-1, 1]^3$. The Radon transform of $\widetilde V$ is well-defined, but not straightforward to compute numerically. We can approximate $\mathcal R \widetilde V(s, \uv n)$ by extracting a discrete 2D slice of $\widetilde V$ and summing its pixels. For that, let $\mathbf u$ and $\mathbf v$ be two vectors orthogonal to $\uv n$. We define the slice ${S_V(s, \uv n, \mathbf u, \mathbf v): [h_S] \times [w_S] \rightarrow \mathbb R}$ by
\begin{equation}
    S_V(s, \uv n, \mathbf u, \mathbf v)(i, j) \coloneqq \widetilde V(s\uv n + i' \mathbf u + j' \mathbf v)
\end{equation}
where  $i' \coloneqq 2\frac{i}{h_S-1}-1$ and  $j' \coloneqq 2\frac{j}{w_S-1}-1$ map integer coordinates $i$ and $j$ to the range $[-1, 1]$. $S_V(s, \uv n, \mathbf u, \mathbf v)$ is a 2D image, which can be  efficiently extracted with the function \texttt{grid\_sample} of PyTorch \cite{pytorch} to be used as input for a neural network. It is the slice of $V$ which is oriented by $\mathbf{u}$ and $\mathbf{v}$, whose normal vector is $\uv{n}$, and which contains the point $s\uv{n}$. The lengths of $\mathbf{u}$ and $\mathbf{v}$ determine the scale of the slice and the angle between them controls the shear (which allows us to conveniently introduce augmentations during training).
For tomographic reconstruction, we choose them perpendicular to each other, with unit length. Therefore, let $\uv n \mapsto \left(\uv u(\uv n), \uv v(\uv n)\right)$ be a function such that $(\uv n, \uv u(\uv n), \uv v(\uv n))$ is an orthonormal basis (see \Cref{sec:about_u_and_v}).
We define the approximate Radon transform of $V$ as
\begin{align}
    \label{eq:radon_transform_approx}
    \mathcal R_\Sigma V(s, \uv n) &\coloneqq \frac{1}{h_S w_S} \sum_{i=0}^{h_S-1}\sum_{j=0}^{w_S-1} S_V(s, \uv n, \uv u(\uv n), \uv v(\uv n))(i, j) \\[-15pt]
    & \approx \mathcal{R}\widetilde{V}(s, \uv n) \nonumber
\end{align}

\subsection{ToNNO}
The idea of this work is to replace the sum of \Cref{eq:radon_transform_approx} with a trained 2D neural network that maps a 2D image to a real number, before applying the inverse Radon transform (see \Cref{fig:main_diagram}). We define
\begin{equation}
    \mathcal R_{g_\theta} V(s, \uv n) \coloneqq g_\theta\left(S_V(s, \uv n, \uv u(\uv n), \uv v(\uv n))\right)
\end{equation}
The output of our method is a volume ${G: [D] \times [H] \times [W] \rightarrow \mathbb R}$ (in practice, to facilitate evaluation, we set the output shape to be equal to the input shape, that is, $D=d$, $H=h$ and $W=w$) such that
\begin{equation}
    G(i, j, k) \approx \mathcal R^{-1}(\mathcal R_{g_\theta} V)(i', j', k')
\end{equation}
where $i' = 2\frac{i}{D-1} -1$, $j' = 2\frac{j}{H-1} -1$ and $k' = 2\frac{k}{W-1} -1$ map integer coordinates to the range $[-1, 1]$.

The integral over $\uv n \in \mathbb S^2$ of \Cref{eq:inv_radon_transform} can be approximated by a sum over a set of L unit vectors that are evenly distributed over the unit sphere $\mathbb S^2$
. To get such a set of unit vectors, we use the Fibonacci lattice: for $\ell \in [L]$, let $y_\ell = 2\frac{\ell}{L-1}-1 \in [-1, 1]$, $r_\ell = \sqrt{1 - y_\ell^2}$, $\theta_\ell = \ell\alpha$ where $\alpha = \pi (3-\sqrt{5})$ is the golden angle, $x_\ell = r_\ell\cos \theta_\ell$ and $z_\ell = r_\ell \sin \theta_\ell$. We then set $\uv n_\ell = (x_\ell, y_\ell, z_\ell)$. We then have
\begin{align}
    & \mathcal R^{-1}(\mathcal R_{g_\theta} V)(i', j', k') \nonumber \\
    \approx \; & \frac{c |\mathbb S^2|}{L}\sum_{l=0}^{L-1} \frac{\partial^2 \mathcal R_{g_\theta} V}{\partial s^2}\left(\left(\begin{smallmatrix}i'\\j'\\k'\end{smallmatrix}\right)  \cdot \uv n_\ell, \uv n_\ell\right)
    \label{eq:inv_radon_transform_discrete}
\end{align}
where $|\mathbb S^2|$ is the surface area of the unit sphere.


It remains to estimate the quantity $\partial^2 \mathcal R_{g_\theta} V(s, \uv n_\ell)/\partial s^2$. In order to do that, for a fixed unit vector $\uv n_\ell$, we sample the function $s \mapsto \mathcal R_{g_\theta} V(s, \uv n_\ell)$ at $M+2$ regular intervals $s_m = 2\tfrac{m}{M-1} -1, m \in \{-1, 0, ..., M\}$, which corresponds to extracting a stack of $M+2$ regularly spaced slices with normal vector $\uv n_\ell$ and feeding each of them to the neural network. Letting $p_m(\uv n_\ell) \coloneqq \mathcal R_{g_\theta}V(s_m, \uv n_\ell)$
, we use finite differences to estimate the second derivative, defining, for $m \in \{0, ..., M-1\}$
\begin{equation}
    q_m(\uv n_\ell) \coloneqq \frac{p_{m+1}(\uv n_\ell) + p_{m-1}(\uv n_\ell) - 2p_m(\uv n_\ell)}{1/(M-1)}
    \label{eq:finite_diff}
\end{equation}
so that 
\begin{equation}
    \label{eq:deriv_approx}
    q_m(\uv n_\ell) \approx \frac{\partial^2 \mathcal R_{g_\theta} V}{\partial s^2}(s_m, \uv n_\ell)
\end{equation}
Now, let $s \mapsto \widetilde q(s, \uv n_\ell)$ be an interpolation of $m \mapsto q_m(\uv n_\ell)$, such that $\widetilde q(-1, \uv n_\ell) = q_0(\uv n_\ell)$, $\widetilde q(1, \uv n_\ell) = q_{M-1}(\uv n_\ell)$ and $\widetilde q(s, \uv n_\ell) = 0$ for $s \notin [-1, 1]$ (which again, can be efficiently computed using the function $\texttt{grid\_sample}$ of PyTorch).
We finally define
\begin{align}
    G(i, j, k) \coloneqq \; & \frac{c |\mathbb S^2|}{L}\sum_{l=0}^{L-1} \widetilde q\left(\left(\begin{smallmatrix}i'\\j'\\k'\end{smallmatrix}\right)  \cdot \uv n_\ell, \uv n_\ell\right)
    \label{eq:inv_radon_transform_final} \\
    \approx \; & \mathcal R^{-1}(\mathcal R_{g_\theta} V)(i', j', k') \nonumber 
\end{align}
The correctness of the implementation can be verified by substituting
$\mathcal R_{g_\theta}$ for $\mathcal R_\Sigma$ (\Cref{eq:radon_transform_approx}) and checking that in this case, $G \approx V$.



\subsection{Averaged CAM and Tomographic CAM}
\label{sec:tomo_cam}

\begin{figure*}[t]
    \centering

    \captionsetup[subfigure]{labelformat=empty}
    \captionsetup[subfigure]{justification=centering}
    \captionsetup[subfigure]{font=figurefont}
    
    \begin{minipage}[c]{0.07\textwidth}
        \subcaption{Averaged}
    \end{minipage}
    \begin{minipage}[c]{0.11\textwidth}
        \includegraphics[width=\textwidth, trim={0 0cm 0 0cm}, clip]{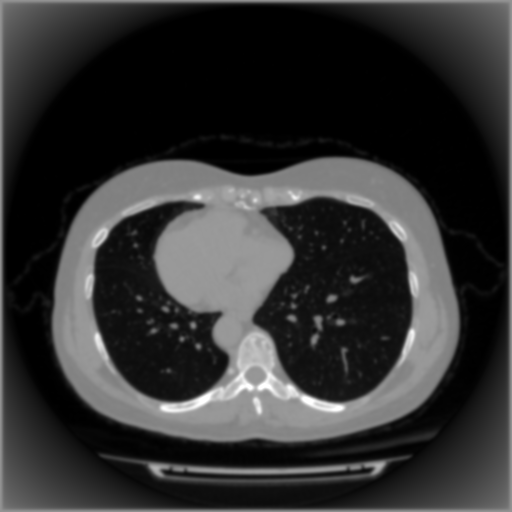}
    \end{minipage}
    \begin{minipage}[c]{0.11\textwidth}
        \includegraphics[width=\textwidth, trim={0 0cm 0 0cm}, clip]{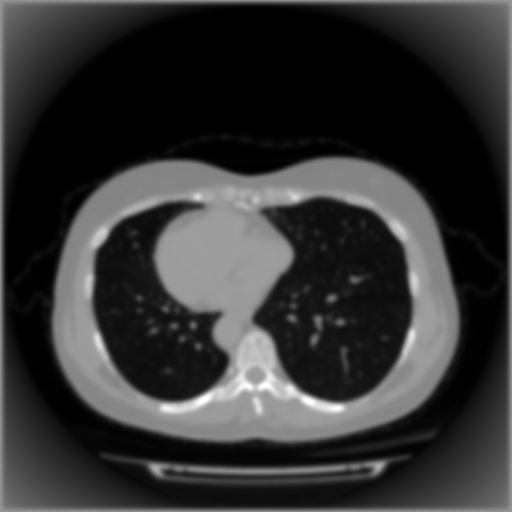}
    \end{minipage}
    \begin{minipage}[c]{0.11\textwidth}
        \includegraphics[width=\textwidth, trim={0 0cm 0 0cm}, clip]{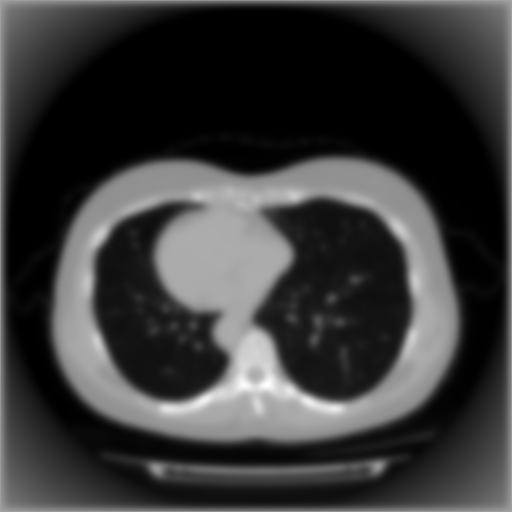}
    \end{minipage}
    \begin{minipage}[c]{0.11\textwidth}
        \includegraphics[width=\textwidth, trim={0 0cm 0 0cm}, clip]{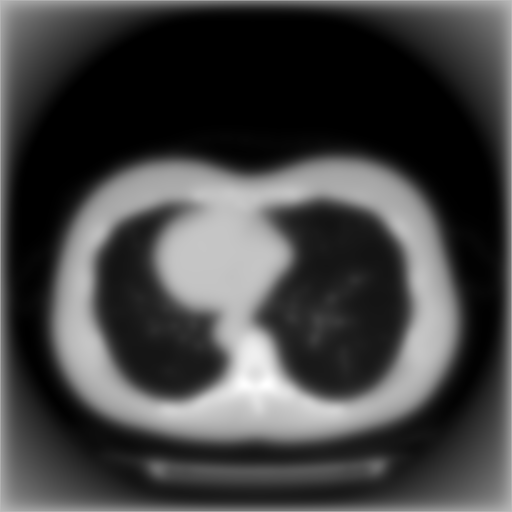}
    \end{minipage}
    \begin{minipage}[c]{0.11\textwidth}
        \includegraphics[width=\textwidth, trim={0 0cm 0 0cm}, clip]{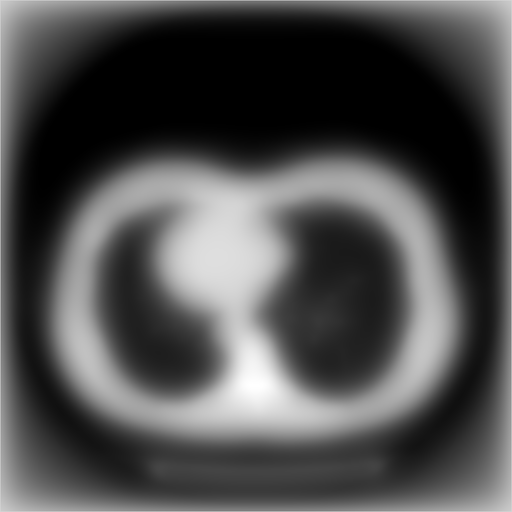}
    \end{minipage}
    \begin{minipage}[c]{0.11\textwidth}
        \includegraphics[width=\textwidth, trim={0 0cm 0 0cm}, clip]{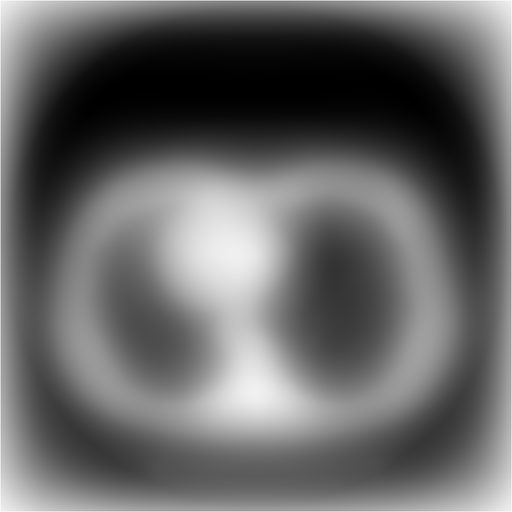}
    \end{minipage}
    \begin{minipage}[c]{0.11\textwidth}
        \includegraphics[width=\textwidth, trim={0 0cm 0 0cm}, clip]{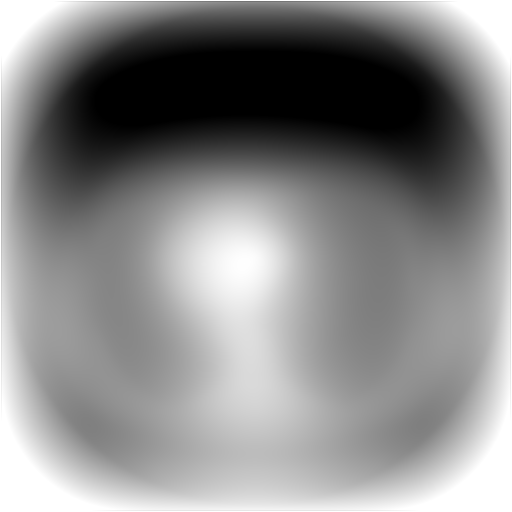}
    \end{minipage}
    \begin{minipage}[c]{0.11\textwidth}
        \includegraphics[width=\textwidth, trim={0 0cm 0 0cm}, clip]{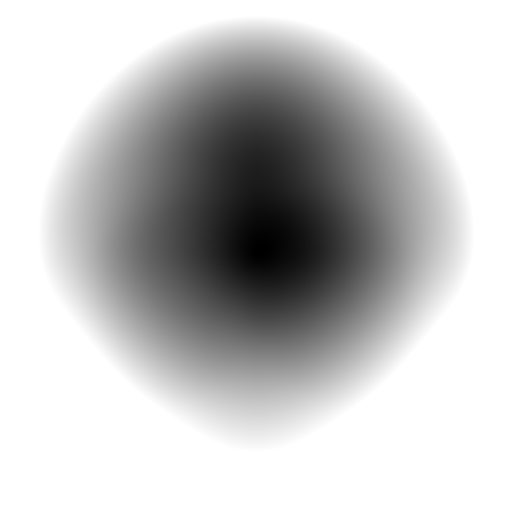}
    \end{minipage}

    \begin{minipage}[c]{0.07\textwidth}
        \subcaption{Tomographic}
    \end{minipage}
    \begin{minipage}[c]{0.11\textwidth}
        \includegraphics[width=\textwidth, trim={0 0cm 0 0cm}, clip]{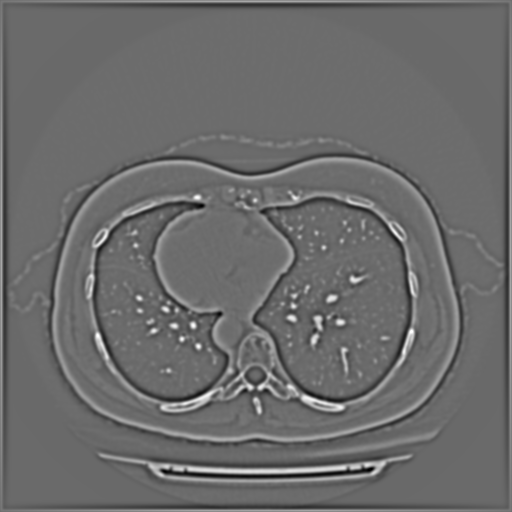}
    \end{minipage}
    \begin{minipage}[c]{0.11\textwidth}
        \includegraphics[width=\textwidth, trim={0 0cm 0 0cm}, clip]{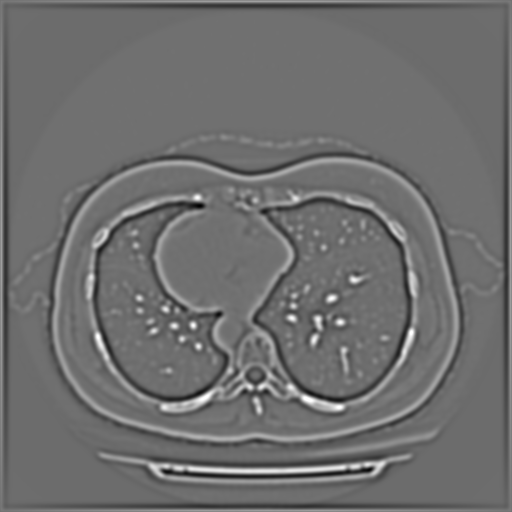}
    \end{minipage}
    \begin{minipage}[c]{0.11\textwidth}
        \includegraphics[width=\textwidth, trim={0 0cm 0 0cm}, clip]{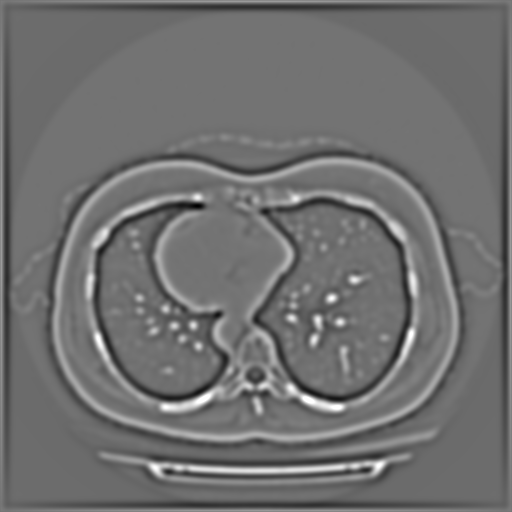}
    \end{minipage}
    \begin{minipage}[c]{0.11\textwidth}
        \includegraphics[width=\textwidth, trim={0 0cm 0 0cm}, clip]{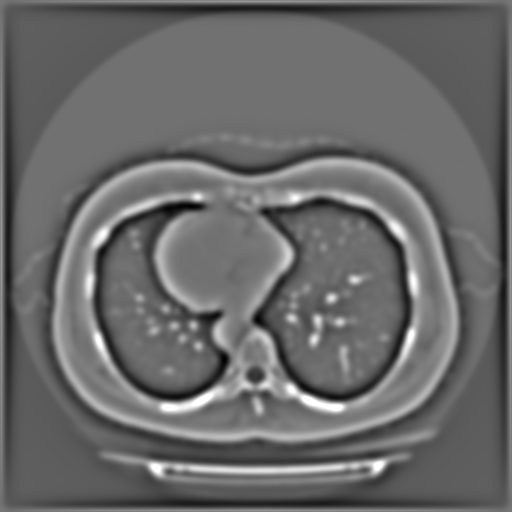}
    \end{minipage}
    \begin{minipage}[c]{0.11\textwidth}
        \includegraphics[width=\textwidth, trim={0 0cm 0 0cm}, clip]{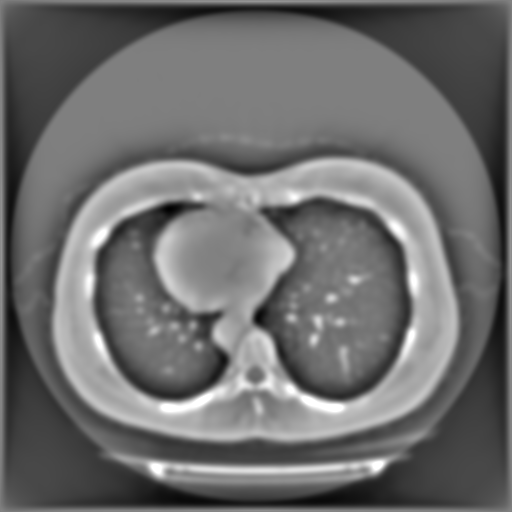}
    \end{minipage}
    \begin{minipage}[c]{0.11\textwidth}
        \includegraphics[width=\textwidth, trim={0 0cm 0 0cm}, clip]{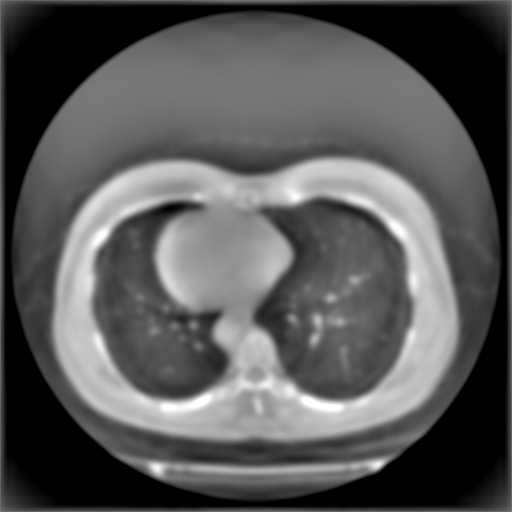}
    \end{minipage}
    \begin{minipage}[c]{0.11\textwidth}
        \includegraphics[width=\textwidth, trim={0 0cm 0 0cm}, clip]{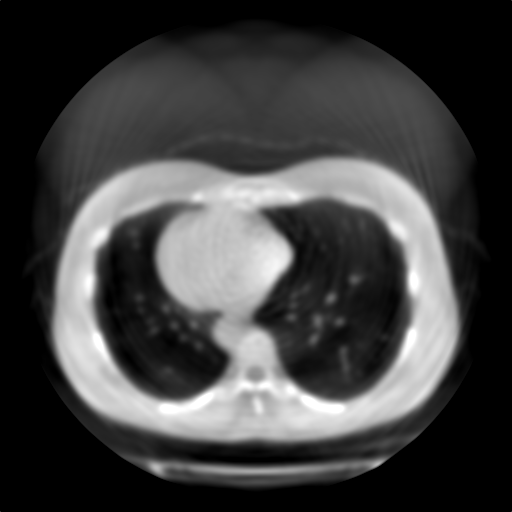}
    \end{minipage}
    \begin{minipage}[c]{0.11\textwidth}
        \includegraphics[width=\textwidth, trim={0 0cm 0 0cm}, clip]{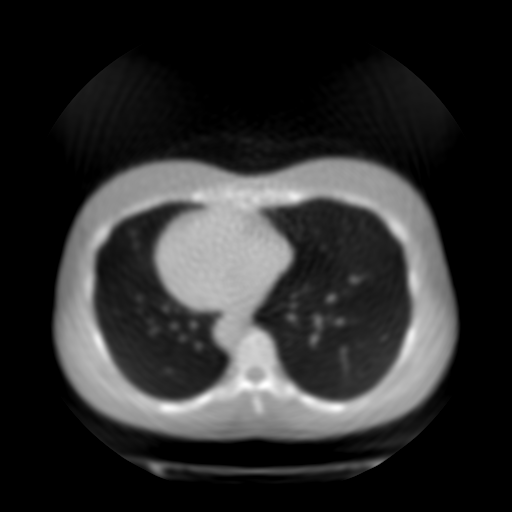}
    \end{minipage}

    \begin{minipage}[t]{0.07\textwidth}\vspace{-0pt}
        \subcaption{}
    \end{minipage}
    \begin{minipage}[t]{0.11\textwidth}\vspace{-0pt}
        \subcaption{Projection size \\ $= 224 \times 224$ }
    \end{minipage}
    \begin{minipage}[t]{0.11\textwidth}\vspace{-0pt}
        \subcaption{Projection size \\ $= 112 \times 112$}
    \end{minipage}
    \begin{minipage}[t]{0.11\textwidth}\vspace{-0pt}
        \subcaption{Projection size \\ $= 56 \times 56$ \\ (equivalent to CAM \\ applied at layer 1)}
    \end{minipage}
    \begin{minipage}[t]{0.11\textwidth}\vspace{-0pt}
        \subcaption{Projection size \\ $= 28 \times 28$ \\ (equivalent to CAM \\ applied at layer 2)}
    \end{minipage}
    \begin{minipage}[t]{0.11\textwidth}\vspace{-0pt}
        \subcaption{Projection size \\ $= 14 \times 14$ \\ (equivalent to CAM \\ applied at layer 3)}
    \end{minipage}
    \begin{minipage}[t]{0.11\textwidth}\vspace{-0pt}
        \subcaption{Projection size \\ $= 7 \times 7$ \\ (equivalent to CAM \\ applied at layer 4)}
    \end{minipage}
    \begin{minipage}[t]{0.11\textwidth}\vspace{-0pt}
        \subcaption{Projection size \\ $= 3 \times 3$}
    \end{minipage}
    \begin{minipage}[t]{0.11\textwidth}\vspace{-0pt}
        \subcaption{Projection size \\ $= 1 \times 1$ \\ ($\mathcal{R}_\Sigma$, \Cref{eq:radon_transform_approx})}
    \end{minipage}

    \caption{\label{fig:input_recon} Input reconstructions for different average pooling kernel sizes.}
\end{figure*}

Class activation mapping (CAM) methods such as GradCAM \cite{selvaraju2017grad} and LayerCAM \cite{jiang2021layercam} are currently the most common method for weakly supervised semantic segmentation. For a given slice, they provide a heatmap of shape $h_{\mathrm{CAM}} \times w_{\mathrm{CAM}}$, examples of which are shown in columns 3 and 5 of \Cref{fig:heatmaps}.
$h_{\mathrm{CAM}}$ and $w_{\mathrm{CAM}}$ correspond to the resolution of the features at the layer from which the class activation maps are extracted. For a ResNet with input shape $224 \times 224$, the resolution at the $i$-th layer is $h_{\mathrm{CAM}} = w_{\mathrm{CAM}} = 224 / 2^{i+1}$. We propose to average these CAM heatmaps across multiple rotations. For a given normal vector $\uv n_\ell$, we extract the $M+2$ slices $S_V(s_m, \uv n_\ell, \uv u_\ell, \uv v_\ell)$ and obtain the heatmap for each slice by applying LayerCAM to the neural network $g_\theta$. We denote by $p_m^{ab}(\uv n_\ell)$ the value of the location $(a, b) \in [h_{\mathrm{CAM}}] \times [w_{\mathrm{CAM}}]$ of the $m$-th heatmap. 
Let $(s, u, v) \mapsto \widetilde p(s, u, v, \uv n_\ell)$ be an interpolation of the 3D volume $(m, a, b) \mapsto p_m^{ab}(\uv n_\ell)$.
A point with coordinates $\begin{smallmatrix}(i' & j' & k')\end{smallmatrix}$ in the canonical coordinate system has coordinates $s = \begin{smallmatrix}(i' & j' & k')\end{smallmatrix} \cdot \uv n_\ell$, $u = \begin{smallmatrix}(i' & j' & k')\end{smallmatrix} \cdot \uv u_\ell$, $v = \begin{smallmatrix}(i' & j' & k')\end{smallmatrix} \cdot \uv v_\ell$ in the coordinate system $(\uv n_\ell, \uv u_\ell, \uv v_\ell)$.
We thus define the average heatmap, that we call \textbf{Averaged CAM}:
\begin{align}
    \label{eq:inv_radon_transform_cam}
    & G_{\mathrm{avgCAM}}(i, j, k) \\ 
    \coloneqq \; & \frac{c |\mathbb S^2|}{L}\sum_{l=0}^{L-1} \widetilde p\left(\left(\begin{smallmatrix}i'\\j'\\k'\end{smallmatrix}\right)  \cdot \uv n_\ell, \left(\begin{smallmatrix}i'\\j'\\k'\end{smallmatrix}\right) \cdot \uv u_\ell, \left(\begin{smallmatrix}i'\\j'\\k'\end{smallmatrix}\right) \cdot \uv v_\ell, \uv n_\ell\right) \nonumber
\end{align}

Using class activation maps from deeper layers (and thus of lower spatial resolution) leads to blurrier results, as can be seen in Figures \labelcref{fig:gvf_layers1to4}, \labelcref{fig:autopet_layers1to4}, \labelcref{fig:mosmed_layers1to4}, and \labelcref{fig:duke_layers1to4} in the Appendix. \Cref{fig:cam_layer_plots} confirms that the performance of Averaged LayerCAM quickly deteriorates when it is applied at deeper layers. We find that we can partially correct this by applying the second derivative along the slice axis separately for each location $(a, b)$, that is, we define $q_m^{ab}$ by replacing $p_m$ by $p_m^{ab}$ in \Cref{eq:finite_diff}. We then define 
\textbf{Tomographic CAM}, by substituting $\widetilde{p}$ with $\widetilde{q}$ (which is the interpolation of $q$, as defined for $p$) in \Cref{eq:inv_radon_transform_cam}:
\begin{align}
    \label{eq:tomo_cam}
    & G'_{\mathrm{avgCAM}}(i, j, k) \\ 
    \coloneqq \; & \frac{c |\mathbb S^2|}{L}\sum_{l=0}^{L-1} \widetilde q\left(\left(\begin{smallmatrix}i'\\j'\\k'\end{smallmatrix}\right)  \cdot \uv n_\ell, \left(\begin{smallmatrix}i'\\j'\\k'\end{smallmatrix}\right) \cdot \uv u_\ell, \left(\begin{smallmatrix}i'\\j'\\k'\end{smallmatrix}\right) \cdot \uv v_\ell, \uv n_\ell\right) \nonumber
\end{align}
\Cref{fig:cam_layer_plots} shows that Tomographic LayerCAM indeed allows to maintain good performance for deeper layers.

\begin{figure}[htb]
  \centering
   \includegraphics[width=1\linewidth]{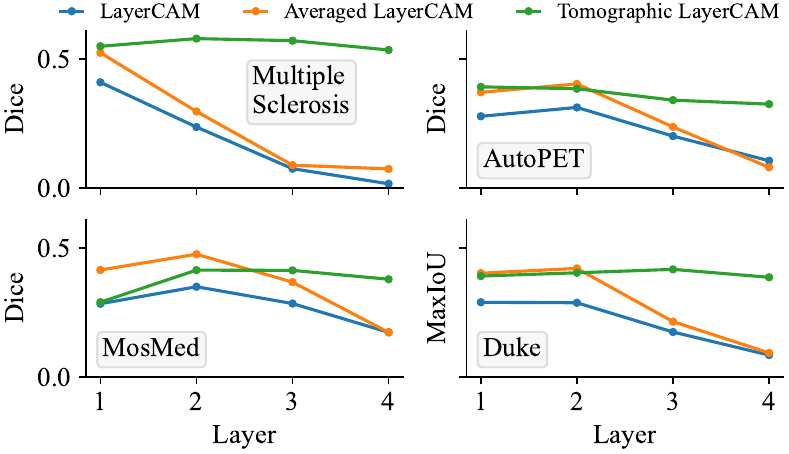} \\
   \caption{\textbf{Dice score for different layers} of LayerCAM, Averaged LayerCAM and Tomographic LayerCAM.}
   \label{fig:cam_layer_plots}
\end{figure}

For gaining further insight into this, we can assume that the value at position $(a, b) \in [h_{\mathrm{CAM}}] \times [w_{\mathrm{CAM}}]$ roughly represents the amount of activation for a region of shape $224 / h_{\mathrm{CAM}} \times 224 / w_{\mathrm{CAM}}$ centered on the pixel ${(a \cdot 224/h_{\mathrm{CAM}},  b \cdot 224/w_{\mathrm{CAM}})}$ in the input image. Thus, we can obtain a "CAM" version of $\mathcal{R}_\Sigma$ of \Cref{eq:radon_transform_approx} by average-pooling the input image with a kernel size and stride of $(224/h_{\mathrm{CAM}}, 224/w_{\mathrm{CAM}})$. We can then use the formulas of Averaged CAM and Tomographic CAM (Equations (\ref{eq:inv_radon_transform_cam}) and (\ref{eq:tomo_cam})) to reconstruct the input image. The results are presented on \Cref{fig:input_recon}. As can be seen, when the CAM maps have high resolution, then simple averaging does allow to recover the input. But when the CAM maps have lower resolution, incorporating the second derivative becomes necessary. In between, the "Averaged" reconstructions are blurry, and the "Tomographic" reconstructions (i.e., with second derivative) are too sharp, as if a filter had been applied on the image. Future work could look into a way of reconstructing these intermediate cases, for example using iterative reconstruction techniques.

\subsection{Hyperparameters}

\paragraph{Tomographic reconstruction} We experiment with different numbers of slices (M in \Cref{eq:finite_diff}) and angles (L in \Cref{eq:inv_radon_transform_discrete}). Reconstructions of one sample of the MosMed dataset are shown in \Cref{fig:l_and_k} of the Appendix for $(M, L) \in \{25, 50, 100, 200\} \times \{500, 1000, 2000, 5000\}$. Increasing the number of slices $M$ increases the sharpness of the reconstruction but also increases the amount of noise. The number of angles $L$ reduces the noise. In our experiments, we choose $M = 100$ and $L = 2000$ and we set the output shape to be equal to the shape of the input volume. For Averaged CAM and Tomographic CAM, we set $L = 1000$ because obtaining the class activation maps involves more computations. All interpolations are performed with trilinear interpolation. With these parameters, the reconstruction time is around 40 seconds for ToNNO and 1 minute for Averaged and Tomographic CAM on a NVIDIA GeForce RTX 3090 GPU. This is slower than slice-wise CAM methods, which take less than a second per volume on a GPU.

\paragraph{Classifier choice}

The neural network should have three characteristics: fast inference (as many inferences are needed for tomographic reconstruction), availability of pre-trained weights, and absence of batch normalization (BN) layers \cite{ioffe2015batch}. We hypothesize that the latter is important because BN breaks the independence between samples within a batch, leading to a low-loss but less informative solution, as reported in \cite{he2020momentum, cpc}. For example, if one slice is identified by the classifier as containing a discriminative region, then intra-batch information leakage caused by BN could enable the classifier to also output high logits for slices of the volume that do not contain any discriminative region, leading to a lower loss but useless predictions.
In the end, we select two ResNets \cite{resnet, wightman2021resnet} from the {\bf timm} \cite{timm2022} library, for which ImageNet \cite{deng2009imagenet} pretrained weights are available:
\begin{itemize}
    \item The \textbf{ResNet-10-T} \cite{wightman2021resnet} which is a lightweight ResNet. It comes with batch normalization layers, so we explore two modifications: freezing them or replacing them with group normalization layers \cite{group_norm} (still initializing the other layers with the pretrained weights).
    \item The \textbf{ResNet-50} \cite{resnet} with group normalization. Because of heavier overfitting, we train it for only 40\% of the number of iterations of the ResNet-10-T. 
\end{itemize}
We experimented with more recent backbones, such as EfficientNet \cite{efficientnet}, ConvNeXt \cite{convnext}, ViT \cite{vit} and Swin \cite{swin}, but were unable to surpass the more simple ResNets. Careful tuning of training hyper-parameters may allow them to achieve better performance but this is not the focus of our work.

Our results in \Cref{tab:model_results} show that the ResNet-10-T with group normalization and frozen batch normalization are on par, and better than with vanilla batch normalization. We did not find the ResNet-50 to be significantly better than the ResNet-10-T.
The ResNet-10-T with random initialization achieves much lower performance, confirming the benefit of using pretrained weights. In all the other experiments presented in this paper, we use the ResNet-10-T with frozen batch normalization and ImageNet pretrained weights.

\begin{table}[ht]
    \setlength{\tabcolsep}{4pt}
    \centering
    
    \begin{tabular}{lccc}

   {\bf   Model }& \bf{F1-score} & \bf{Dice/IoU} & {\bf BA} \\
    \toprule
        R-10-T, batch norm & 0.48 & 0.31 & 0.84 \\
         R-10-T, frozen batch norm & 0.55 & 0.39 & 0.84 \\
         R-10-T, group norm & 0.54 & 0.36 & 0.85 \\
         R-10-T, not pretrained & 0.28 & 0.27 & 0.68 \\
         R-50, group norm & 0.56 & 0.40 & 0.84 \\
    \bottomrule
    \end{tabular}
    \caption{\textbf{Comparison of different model configurations}. The metrics were averaged over the four datasets. Per-dataset results are presented in \Cref{sec:detailed_model_results}.}
    \label{tab:model_results}
\end{table}

%% file: 04_experiments.tex
\section{Experiments}
\label{sec:experiments}

\subsection{Datasets}
We use 4 different datasets for our experiments:

\begin{itemize}
    \item \textbf{Multiple Sclerosis (MS) (brain MRI)} This large private dataset consists of 9113 studies.
The goal is to segment gadolinium-enhancing lesions, which are typically hyperintense on the post-gadolinium injection T1 weighted scan. This abnormal gadolinium uptake is due to blood-brain barrier breakdown and is a sign of acute multiple sclerosis activity.

    \item \textbf{AutoPET-II (whole body PET-CT)} This public dataset \cite{autopet} consists of 1,014 FDG-PET/CT pairs. The objective of this exam is to segment FDG-avid tumours.

    \item \textbf{MosMedData COVID-19  (thoracic CT)} This public dataset \cite{morozov2020mosmeddata}  consists of 1110 studies of thoracic CT scans. The goal is to segment pulmonary COVID-19 lesions.

    \item \textbf{Duke breast cancer MRI} This public dataset \cite{duke} consists of breast MRIs of 922 biopsy-confirmed invasive breast cancer patients. The goal is to detect the tumours.

\end{itemize}
We split each dataset into training and validation sets at the patient level. Dataset characteristics are presented in \Cref{sec:dataset_characteristics} and preprocessing details are explained in \Cref{sec:preproc}.
For the Multiple Sclerosis and AutoPET datasets, ground truth segmentation masks are available for all samples. Patients with non-empty segmentation mask are considered positive while the rest are considered negative. For the MosMed dataset, ground truth segmentation masks are available for 50 patients, which we all put in the validation set. The positive/negative labels of all patient are available separately. For the Duke dataset, all patients are positive for breast cancer, and exactly one bounding box per patient, delimiting the main tumour, is available. Our method needs both positive and negative samples, so we separate the left and right breast of each patient and consider the breast with bounding box to be positive.

\begin{figure*}[htb]
    
    \centering
    \includegraphics[width=1\textwidth, trim={0 1.3cm 0 2cm}, clip]{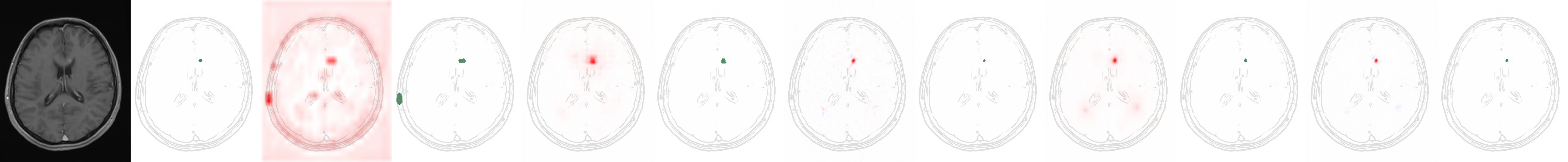} \\
    \includegraphics[width=1\textwidth, trim={0 10cm 0 5cm}, clip]{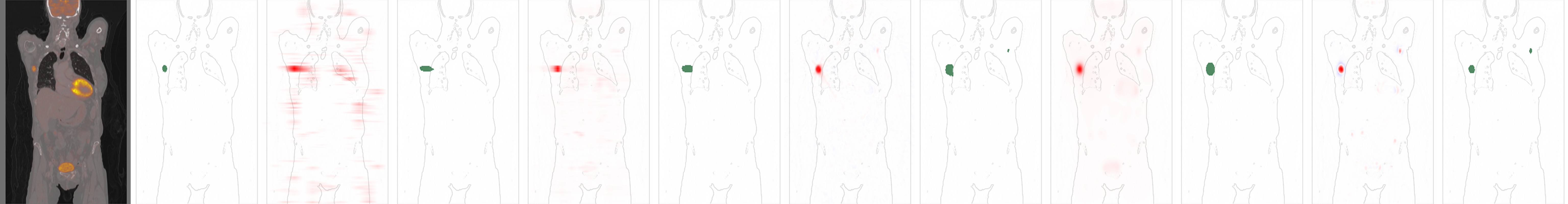} \\
    \includegraphics[width=1\textwidth, trim={0 3cm 0 3cm}, clip]{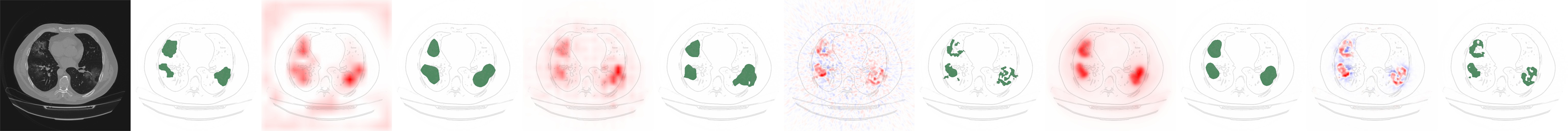} \\
    \includegraphics[width=1\textwidth, trim={0 8cm 0 6cm}, clip]{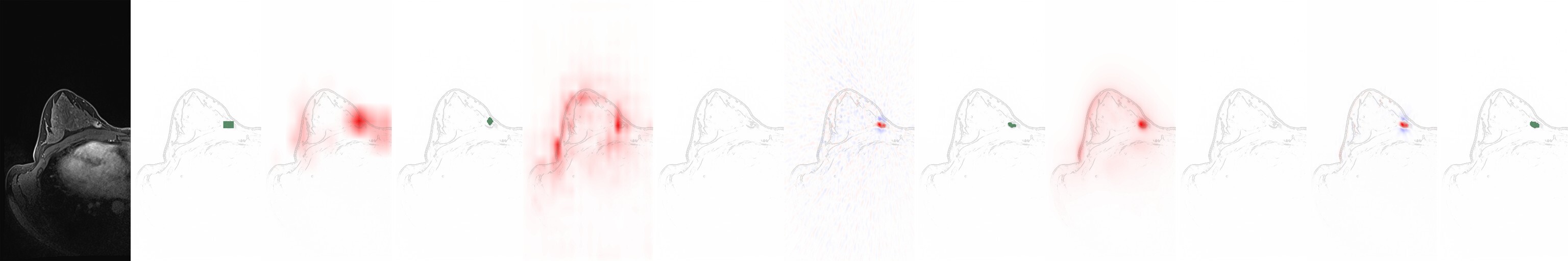} \\
    
    \captionsetup[subfigure]{labelformat=empty}
    \captionsetup[subfigure]{justification=centering}
    \captionsetup[subfigure]{font=figurefont}

    \hspace{-3pt}
    \begin{minipage}[t]{0.079\textwidth}\vspace{-10pt}
    \subcaption{Input}
    \end{minipage}
    \begin{minipage}[t]{0.079\textwidth}\vspace{-10pt}
    \subcaption{Ground truth}
    \end{minipage}
    \begin{minipage}[t]{0.079\textwidth}\vspace{-10pt}
    \subcaption{GradCAM}
    \end{minipage}
    \begin{minipage}[t]{0.079\textwidth}\vspace{-10pt}
    \subcaption{GradCAM\\binarized}
    \end{minipage}
    \begin{minipage}[t]{0.079\textwidth}\vspace{-10pt}
    \subcaption{LayerCAM}
    \end{minipage}
    \begin{minipage}[t]{0.079\textwidth}\vspace{-10pt}
    \subcaption{LayerCAM\\binarized}
    \end{minipage}
    \begin{minipage}[t]{0.079\textwidth}\vspace{-10pt}
    \subcaption{ToNNO}
    \end{minipage}
    \begin{minipage}[t]{0.079\textwidth}\vspace{-10pt}
    \subcaption{ToNNO binarized}
    \end{minipage}
    \begin{minipage}[t]{0.079\textwidth}\vspace{-10pt}
    \subcaption{Averaged LayerCAM}
    \end{minipage}
    \begin{minipage}[t]{0.079\textwidth}\vspace{-10pt}
    \subcaption{Averaged LayerCAM binarized}
    \end{minipage}
    \begin{minipage}[t]{0.079\textwidth}\vspace{-10pt}
    \subcaption{Tomographic LayerCAM}
    \end{minipage}
    \begin{minipage}[t]{0.079\textwidth}\vspace{-10pt}
    \subcaption{Tomographic LayerCAM binarized}
    \end{minipage}
    
    \caption{\label{fig:heatmaps} \textbf{Examples of heatmaps and segmentations obtained with each method.} From top to bottom: Multiple Sclerosis, AutoPET, MosMed and Duke datasets. The images were cropped. For GradCAM, LayerCAM, Averaged LayerCAM and Tomographic LayerCAM, we chose the layer which gave the best results. The binarized heatmaps are obtained using the process described in the \Cref{sec:binarization_and_metrics}. In the heatmaps, blue represents negative values and red represents positive values. More examples are provided in the Appendix.}
\end{figure*}

\subsection{Training procedure}
During a training step, we sample a batch of $B$ samples $\left\{(V_{i_1}, y_{i_1}), ..., (V_{i_B}, y_{i_B})\right\}$ with replacement from our training dataset. We then extract $M_\mathrm{train}$ randomly oriented slices of shape $224 \times 224$ from each volume and apply augmentations. We concatenate the $B \times M_\mathrm{train}$ slices in a batch and perform a training step by associating to each slice the label of the volume from which it came. More details can be found in \Cref{sec:detailed_training_procedure}.

\subsection{Baselines}
We compare our method to class activation methods, namely GradCAM and LayerCAM. We apply them to the ResNet-10-T with frozen batch normalization at the output of layers 1, 2, 3 and 4, to each axial slice of the input volume, and obtain the output volume by stacking the class activation maps back along the axial axis. To be input to the classifier, each slice is resampled to match the input shape of the neural network ($224 \times 224$), and the resulting class activation map is upsampled to match the original slice shape. The output volume thus has the same shape as the input volume.

In \cref{sec:rebuttal_additions}, we also provide results for GradCAM++ \cite{gradcam_pp} and ScoreCAM \cite{score_cam}. Furthermore, we tried to apply the CAM methods along the 3 spatial axes, averaging the outputs. In some cases, this resulted in significant improvements (\cref{sec:rebuttal_additions}).

\subsection{Evaluation}

\label{sec:eval}

In order to compute segmentation metrics, we first binarize the heatmaps using the process described in \Cref{sec:binarization_and_metrics}. We then compute the connected components of the binarized heatmaps and the ground truth segmentation and report precision, recall and F1-score. We also report the dice score. For the Duke dataset, as no ground truth segmentation masks are available (we only have one bounding box annotation per patient), we replace the dice score with the intersection-over-union of the predicted bounding box which best matches the ground truth bounding box. In the different tables, this metric is called MaxIoU. As there is at most one ground truth bounding box per patient, despite there sometimes being multiple tumours, many predicted connected components that are considered as false positives may actually be true positives. Thus, precision and F1-score are less relevant for this dataset.

\begin{table*}[t]
    \centering
    \begin{adjustbox}{width={\textwidth},totalheight={\textheight},keepaspectratio}
    \begin{tabular}{lllllll}
    {\bf Dataset} & {\bf Method} & {\bf Precision} & {\bf Recall} & {\bf F1-score} & {\bf Dice/MaxIoU} & {\bf Balanced accuracy}  \\
    \toprule
    \multirow{4}{*}{\shortstack[l]{Multiple\\Sclerosis}}
        & LayerCAM (layer 1) & 0.76 & 0.78 & 0.73 & 0.41 & 0.89 \\
        & ToNNO & 0.82 (\textcolor[HTML]{4c9e62}{+0.06})& \textbf{0.88 (\textcolor[HTML]{4c9e62}{+0.10})} & 0.82 (\textcolor[HTML]{4c9e62}{+0.09})& 0.53 (\textcolor[HTML]{4c9e62}{+0.12})& 0.90 (\textcolor[HTML]{4c9e62}{+0.01})\\
        & Averaged LayerCAM (layer 1) & 0.86 (\textcolor[HTML]{4c9e62}{+0.10})& 0.87 (\textcolor[HTML]{4c9e62}{+0.09})& \textbf{0.84 (\textcolor[HTML]{4c9e62}{+0.11})} & 0.52 (\textcolor[HTML]{4c9e62}{+0.11})& 0.91 (\textcolor[HTML]{4c9e62}{+0.02})\\
        & Tomographic LayerCAM (layer 2) & \textbf{0.87 (\textcolor[HTML]{4c9e62}{+0.11})} & 0.87 (\textcolor[HTML]{4c9e62}{+0.09})& \textbf{0.84 (\textcolor[HTML]{4c9e62}{+0.11})} & \textbf{0.58 (\textcolor[HTML]{4c9e62}{+0.17})} & \textbf{0.94 (\textcolor[HTML]{4c9e62}{+0.05})} \\
    \midrule
    \multirow{4}{*}{\shortstack[l]{AutoPET}}
        & LayerCAM (layer 2) & 0.59 & 0.29 & 0.35 & 0.31 & 0.80 \\
        & ToNNO & 0.52 (\textcolor{BrickRed}{-0.07})& 0.41 (\textcolor[HTML]{4c9e62}{+0.12})& 0.39 (\textcolor[HTML]{4c9e62}{+0.04})& 0.32 (\textcolor[HTML]{4c9e62}{+0.01})& 0.74 (\textcolor{BrickRed}{-0.06})\\
        & Averaged LayerCAM (layer 2) & 0.65 (\textcolor[HTML]{4c9e62}{+0.06})& 0.33 (\textcolor[HTML]{4c9e62}{+0.04})& 0.40 (\textcolor[HTML]{4c9e62}{+0.05})& \textbf{0.40 (\textcolor[HTML]{4c9e62}{+0.09})} & \textbf{0.83 (\textcolor[HTML]{4c9e62}{+0.03})} \\
        & Tomographic LayerCAM (layer 1) & \textbf{0.68 (\textcolor[HTML]{4c9e62}{+0.09})} & \textbf{0.47 (\textcolor[HTML]{4c9e62}{+0.18})} & \textbf{0.49 (\textcolor[HTML]{4c9e62}{+0.14})} & 0.39 (\textcolor[HTML]{4c9e62}{+0.08})& 0.74 (\textcolor{BrickRed}{-0.06})\\
    \midrule
    \multirow{4}{*}{\shortstack[l]{MosMed}}
        & LayerCAM (layer 2) & 0.68 & 0.32 & 0.39 & 0.35 & 0.90 \\
        & ToNNO & 0.69 (\textcolor[HTML]{4c9e62}{+0.01})& 0.45 (\textcolor[HTML]{4c9e62}{+0.13})& 0.50 (\textcolor[HTML]{4c9e62}{+0.11})& 0.35 & 0.93 (\textcolor[HTML]{4c9e62}{+0.03})\\
        & Averaged LayerCAM (layer 2) & \textbf{0.72 (\textcolor[HTML]{4c9e62}{+0.04})} & 0.49 (\textcolor[HTML]{4c9e62}{+0.17})& 0.53 (\textcolor[HTML]{4c9e62}{+0.14})& \textbf{0.48 (\textcolor[HTML]{4c9e62}{+0.13})} & \textbf{0.95 (\textcolor[HTML]{4c9e62}{+0.05})} \\
        & Tomographic LayerCAM (layer 3) & \textbf{0.72 (\textcolor[HTML]{4c9e62}{+0.04})} & \textbf{0.52 (\textcolor[HTML]{4c9e62}{+0.20})} & \textbf{0.55 (\textcolor[HTML]{4c9e62}{+0.16})} & 0.41 (\textcolor[HTML]{4c9e62}{+0.06})& 0.89 (\textcolor{BrickRed}{-0.01})\\
    \midrule
    \multirow{4}{*}{\shortstack[l]{Duke}}
        & LayerCAM (layer 2) & \textcolor{black}{0.17 }& 0.81 & \textcolor{black}{0.24 }& 0.29 & 0.58 \\
        & ToNNO & \textcolor{black}{\textbf{0.43 (\textcolor[HTML]{4c9e62}{+0.26})} }& 0.77 (\textcolor{BrickRed}{-0.04})& \textcolor{black}{\textbf{0.51 (\textcolor[HTML]{4c9e62}{+0.27})} }& 0.37 (\textcolor[HTML]{4c9e62}{+0.08})& \textbf{0.79 (\textcolor[HTML]{4c9e62}{+0.21})} \\
        & Averaged LayerCAM (layer 2) & \textcolor{black}{0.37 (\textcolor[HTML]{4c9e62}{+0.20})}& \textbf{0.84 (\textcolor[HTML]{4c9e62}{+0.03})} & \textcolor{black}{0.47 (\textcolor[HTML]{4c9e62}{+0.23})}& \textbf{0.42 (\textcolor[HTML]{4c9e62}{+0.13})} & 0.75 (\textcolor[HTML]{4c9e62}{+0.17})\\
        & Tomographic LayerCAM (layer 3) & \textcolor{black}{0.41 (\textcolor[HTML]{4c9e62}{+0.24})}& 0.81 (=) & \textcolor{black}{\textbf{0.51 (\textcolor[HTML]{4c9e62}{+0.27})} }& \textbf{0.42 (\textcolor[HTML]{4c9e62}{+0.13})} & \textbf{0.79 (\textcolor[HTML]{4c9e62}{+0.21})} \\
    \midrule

    \end{tabular}
    \end{adjustbox}
    \caption{\textbf{Main quantitative results of our method and best performing baseline method (LayerCAM).} For LayerCAM, Averaged LayerCAM and Tomographic LayerCAM, we selected the layer which provided the best results. Detailed results for each layer are provided in the Appendix. The numbers in color indicate the difference with LayerCAM.
    }
    \label{tab:main_results}
\end{table*}

\subsection{Results}

Main results are summarized in \Cref{tab:main_results}, and results for each method applied at each layer can be found in the Appendix. ToNNO outperforms LayerCAM (which largely outperforms GradCAM as can be seen in the Appendix) in most cases. Averaged LayerCAM and Tomographic LayerCAM achieve even better results.
Tomographic LayerCAM achieves the best results in terms of F1-score, with an average improvement of 0.14 over LayerCAM. In terms of recall, Tomographic LayerCAM achieves particularly large gains over LayerCAM for the AutoPET and MosMed datasets.

All methods achieve their best results on the Multiple Sclerosis dataset. This could be attributed to several factors: the training dataset is an order of magnitude larger than the other datasets and the shape of the lesions is simple.
Our method can be compared to the results presented in \cite{ms_pointwise}, which uses point-wise lesion annotations while we use only image-level labels. Although they do not use the same dataset and the same evaluation process, the results that we achieve are within the same order of magnitude as theirs (F1-score of 0.86 for their method versus 0.84 for Tomographic LayerCAM).

Examples of heatmaps and associated binary segmentations for each method and dataset are shown in \Cref{fig:heatmaps}, and examples of heatmaps for each layer can be found in the Appendix. ToNNO and Tomographic LayerCAM provide much sharper heatmaps than baseline 2D CAM methods and Averaged CAM. While this is an advantage for segmenting the small lesions of the Multiple Sclerosis dataset, it results in undersegmentation of the large lesions of the MosMed dataset.


%% file: 10_conclusion.tex
\section{Conclusion and future work}
\label{sec:conclusion}

In this work, we have presented ToNNO, a novel method that allows to perform dense prediction tasks on 3D images using a 2D encoder. We have shown that in the case where the encoder is trained on a weakly supervised classification task, our method can achieve better results than state-of-the-art CAM methods. We have also proposed Averaged and Tomographic CAM, which allow to integrate CAM methods into our framework in order to achieve even better results.

We believe that ToNNO could be applied in other ways. For example, it could allow to obtain dense features for 3D images using an encoder that is pretrained either on large scale computer vision datasets such as ImageNet \cite{deng2009imagenet}, or by using self-supervised learning techniques directly on slices of 3D medical image datasets.
ToNNO, contrary to CAM methods, could also be used to obtain quantitative heatmaps in the case where the encoder is pretrained in a quantitative way, for example on a regression task.

Future work could concentrate on finding a canonical way---for example using iterative methods---to perform tomographic reconstruction with class activation maps, as we have seen that neither Tomographic CAM nor Averaged CAM are perfect generalizations of ToNNO.







%% file: 12_appendix.tex
\section{Appendix}
\hypersetup{colorlinks=true, linkcolor=black}

\titlecontents{subsection}[3em] {} {\contentslabel{3em}} {} {\titlerule*[1pc]{.}\contentspage}
\DoToC
\hypersetup{colorlinks=true, linkcolor=red}

\vspace{1em}

\subsection{About \texorpdfstring{$\uv u(\uv n)$}{u(n)} and \texorpdfstring{$\uv v(\uv n)$}{v(n)}}
\label{sec:about_u_and_v}

\begin{figure}[htb] 
   \begin{subfigure}{0.335\linewidth}
       \includegraphics[width=\linewidth, trim={0 0cm 0 0cm}, clip]{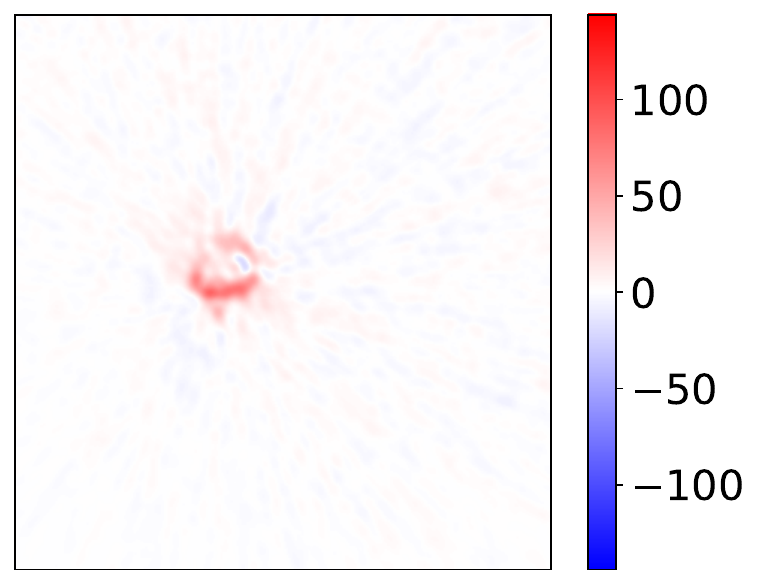}
       \caption{Seed = 0 \hphantom{25pt}}
   \end{subfigure}
\hfill 
   \begin{subfigure}{0.335\linewidth}
       \includegraphics[width=\linewidth, trim={0 0cm 0 0cm}, clip]{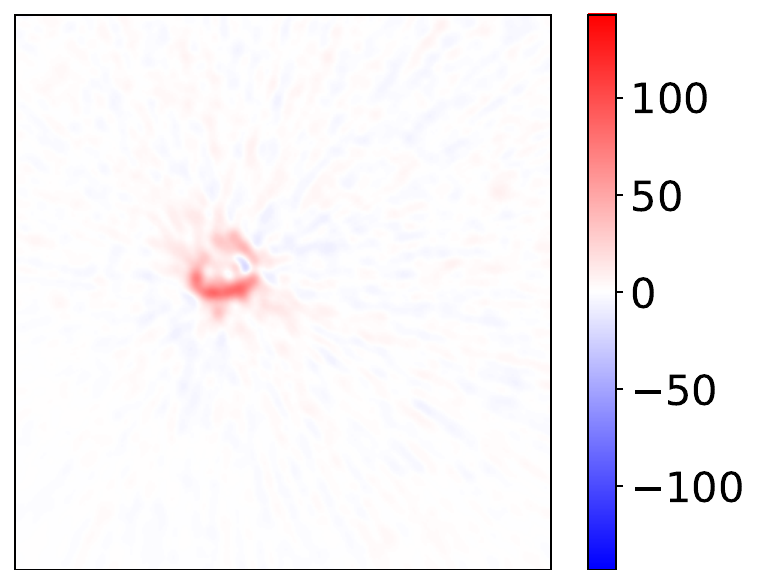}
       \caption{Seed = 1 \hphantom{25pt}}
   \end{subfigure}
\hfill 
   \begin{subfigure}{0.31\linewidth}
       \includegraphics[width=\linewidth, trim={0 0cm 0 0cm}, clip]{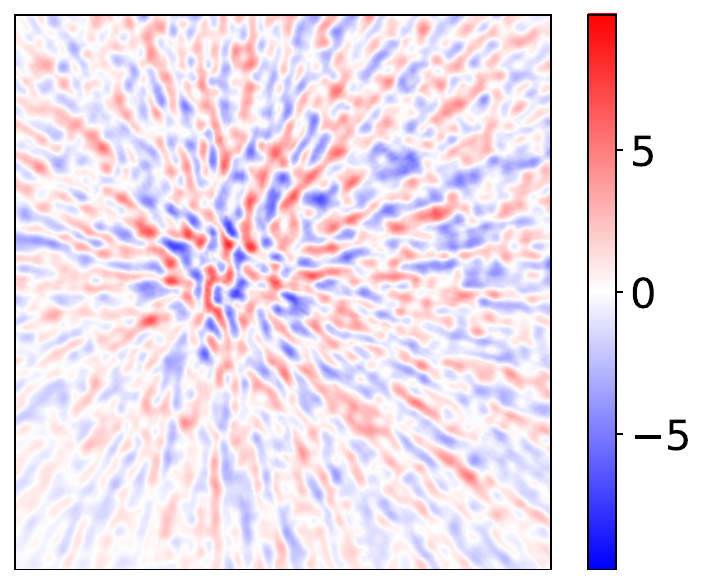}
       \caption{Difference \hphantom{25pt}}
   \end{subfigure}

   \caption{Reconstructions for two different random seeds for a sample of the MosMed dataset.}
   \label{fig:impact_of_seed}
\end{figure}

$\mathcal{R}_\Sigma$, $\mathcal{R}_{g_\theta}$, and thus the output of our method, are dependant on the choice of $\uv u(\uv n)$ and $\uv v(\uv n)$.
During tomographic reconstruction, whenever extracting a slice, we randomly choose $\uv u(\uv n)$ in $\uv n^\perp$ and $\uv v(\uv n)$ in $\uv u(\uv n)^\perp$. Theoretically, $G$ (\Cref{eq:inv_radon_transform_final}) is therefore a random variable. But we find that in practice, the impact of changing the seed is minimal, as shown in \Cref{fig:impact_of_seed}. This could be attributed to the large number of angles $L$ that are used for reconstruction and to ${g_\theta}$ probably being robust to rotations.

\subsection{Impact of the number of slices and angles}

\begin{figure}[h]
  \centering
   \includegraphics[width=1\linewidth]{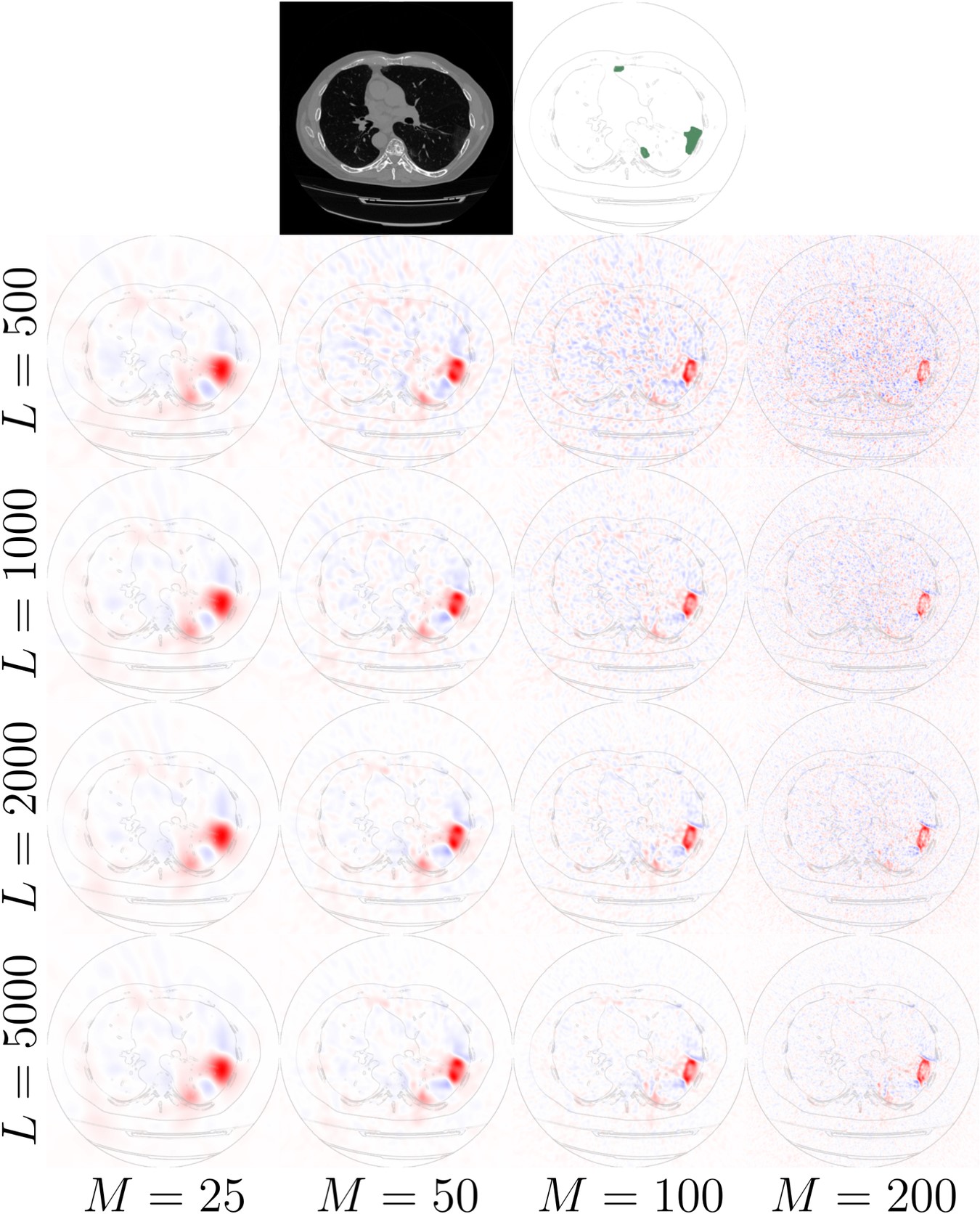} \\
   \caption{\textbf{ToNNO's output for different values of $M$ and $L$} from \Cref{eq:finite_diff,eq:inv_radon_transform_discrete} for a sample of the MosMed dataset. The input image and ground truth segmentation are shown on top.}
   \label{fig:l_and_k}
\end{figure}

\Cref{fig:l_and_k} shows the impact of the number of angles $L$ and the number of slices $M$ on the reconstructions that are produced.

\subsection{Dataset characteristics}
\label{sec:dataset_characteristics}

\Cref{tab:datasets} shows the number of samples used for training and validation in each dataset.

\begin{table}[htb]
  \centering
  \begin{tabular}{@{}lccccc@{}}

\multirow{2}{*}{\bf Dataset} & \multirow{2}{*}{\bf Modality} & \multicolumn{2}{l}{\bf Training} & \multicolumn{2}{l}{\bf Validation} \\
 \multicolumn{2}{c}{} &  \bf pos. & \bf neg. & \bf pos. & \bf neg. \\
 \midrule
    MS  & MRI & 1786 & 6363 & 181 & 783\\
    AutoPET \cite{autopet} & CT, PET & 444 & 457 & 57 & 55\\
    Duke \cite{duke} & MRI & 975 & 699& 92 & 66\\
    MosMed \cite{morozov2020mosmeddata} & CT & 634 & 204& 50 & 50\\
    \bottomrule
  \end{tabular}
  \caption{Dataset characteristics and number of positive and negative samples used for training and validation.}
  \label{tab:datasets}
\end{table}

\subsection{Dataset preprocessing details}
\label{sec:preproc}

\paragraph{Multiple Sclerosis} In this longitudinal dataset, multiple studies (visits) are available for each patient. Each study includes T1, proton-density, T2, gadolinium-enhanced FLAIR and T1 weighted sequences, which were all rigidly registered to a common atlas and cropped. The resulting image shape is $54 \times 222 \times 179$. For a given study, we stack all 5 sequences along the channel axis to be input to the models, and apply channel-wise $z$-normalization. For each study, a ground truth segmentation mask was derived by a consensus of trained experts.

\paragraph{AutoPET} Each study consists of a CT image and an associated Standardized Uptake Value (SUV) image. In order to increase training speed, we cropped all images according to the following protocol: a binary mask was generated by thresholding the SUV image at 0.2. The minimum enclosing bounding box of all the positive voxels was then used to crop the image. The CT scan was finally resampled onto this cropped image. We stack the CT and SUV images along the channel axis. The CT scan is divided by 1000 and the SUV by 10 at the input of the models. Ground truth manual segmentation masks are provided for each study.

\begin{figure}[htb] 
   \begin{subfigure}{0.32\linewidth}
       \includegraphics[width=\linewidth, trim={0 0cm 0 0cm}, clip]{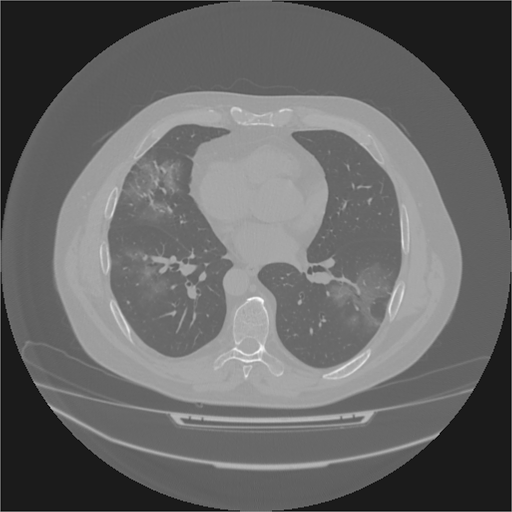}
       \caption{Input}
   \end{subfigure}
\hfill 
   \begin{subfigure}{0.32\linewidth}
       \includegraphics[width=\linewidth, trim={0 0cm 0 0cm}, clip]{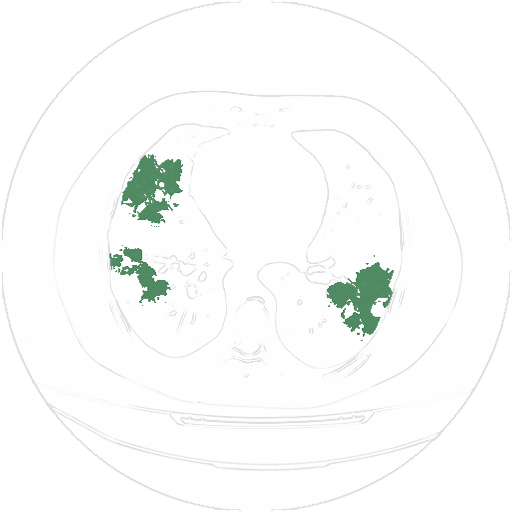}
       \caption{Raw}
   \end{subfigure}
\hfill 
   \begin{subfigure}{0.32\linewidth}
       \includegraphics[width=\linewidth, trim={0 0cm 0 0cm}, clip]{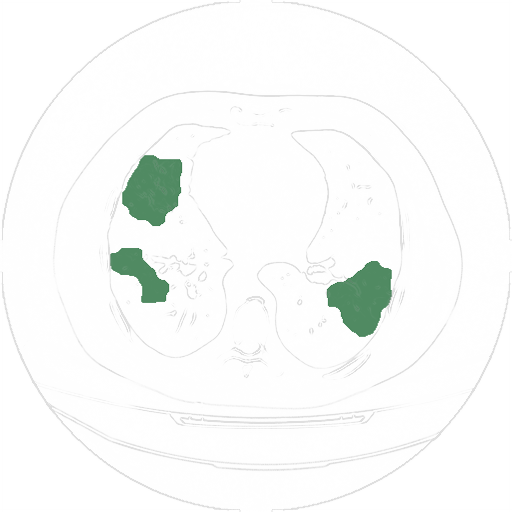}
       \caption{Preprocessed}
   \end{subfigure}

   \caption{The masks of the MosMed datasets were preprocessed in order to remove tiny connected components.}
   \label{fig:mosmed_preproc}
\end{figure}

\paragraph{MosMed} There are 6 classes, named CT-0 to CT-6. CT-0 is the negative class, meaning that no signs of COVID-19 were identified in the scans, and CT-1 to 6 are the positive class and are sorted by increasing order of severity. We use only CT-0 and CT-1. The axial resolution of the CT scans is low, as only every tenth slice was kept in the public release of the dataset. All slices have shape $512 \times 512$. Ground truth semi-manual segmentation masks are available for 50 cases. They contain many tiny connected components, which is not suitable for our evaluation procedure. We therefore preprocessed these masks in order to reduce their number of connected components by applying a binary closing with a sphere of radius 20 followed by a binary opening with a sphere of radius 10. An example of this preprocessing is provided in \Cref{fig:mosmed_preproc}. We divide the volumes by 1000 at the input of the models.

\paragraph{Duke} Each study comprises 6 different acquisitions: a T1-sequence, a fat-saturated pre-contrast T1 sequence, and 4 post-contrast T1 sequences. Trained experts provided bounding boxes delimiting the tumours. There is exactly one bounding-box per patient: even in the case of multiple tumours, only one was annotated. In order to obtain negative volumes, we separated the left and right breasts into two different images: the breast with bounding box was considered positive, the other breast was considered negative, unless the field \textit{Contralateral Breast Involvement} was marked as positive in the clinical data, in which case the other breast was also considered positive. In this work, we only use the pre-contrast and first post-contrast sequences, as these were used by the human annotators to draw the bounding boxes. We stack them along the channel axis and apply channel-wise $z$-normalization. We manually verified that the coordinates of the bounding boxes aligned well with the tumours in all patients.

\subsection{Detailed results for each model.}
\label{sec:detailed_model_results}

\Cref{tab:detailed_model_results} shows results for each model, for each dataset.

\begin{table}[htb]
    \setlength{\tabcolsep}{4pt}
    \centering
    \begin{tabular}{llccc}

   {\bf  Dataset } &{\bf   Model }& \bf{F1-score} & \bf{Dice/IoU} & {\bf BA} \\
    \toprule
    \multirow{5}{*}{\shortstack[l]{Multiple\\Sclerosis}}
        & R-10-T, BN & 0.75 & 0.41 & 0.93 \\
        & R-10-T, FBN & 0.82 & 0.53 & 0.90 \\
        & R-10-T, GN & 0.78 & 0.49 & 0.93 \\
        & R-10-T, GN, NP & 0.21 & 0.20 & 0.52 \\
        & R-50, GN & \textbf{0.83} & \textbf{0.55} & \textbf{0.94} \\
    \midrule
    \multirow{5}{*}{\shortstack[l]{AutoPET}}
        & R-10-T, BN & 0.35 & 0.24 & 0.80 \\
        & R-10-T, FBN & 0.39 & \textbf{0.32} & 0.74 \\
        & R-10-T, GN & 0.36 & 0.27 & 0.77 \\
        & R-10-T, GN, NP & 0.25 & 0.16 & \textbf{0.81} \\
        & R-50, GN & \textbf{0.42} & 0.31 & 0.80 \\
    \midrule
    \multirow{5}{*}{\shortstack[l]{MosMed}}
        & R-10-T, BN & 0.43 & 0.29 & 0.92 \\
        & R-10-T, FBN & 0.50 & 0.35 & \textbf{0.93} \\
        & R-10-T, GN & 0.50 & 0.35 & 0.92 \\
        & R-10-T, GN, NP & 0.39 & 0.32 & 0.83 \\
        & R-50, GN & \textbf{0.54} & \textbf{0.39} & 0.87 \\
    \midrule
    \multirow{5}{*}{\shortstack[l]{Duke}}
        & R-10-T, BN & 0.38 & 0.30 & 0.71 \\
        & R-10-T, FBN & \textbf{0.51} & 0.37 & 0.79 \\
        & R-10-T, GN & 0.50 & 0.35 & \textbf{0.80} \\
        & R-10-T, GN, NP & 0.27 & \textbf{0.39} & 0.54 \\
        & R-50, GN & 0.45 & 0.36 & 0.77 \\
    \bottomrule
    \end{tabular}
    \caption{\textbf{Comparison of different model configurations}. BN: batch normalization, FBN: frozen batch normalization, GN: group normalization, NP: not pretrained.}
    \label{tab:detailed_model_results}
\end{table}

\subsection{Detailed training procedure}

\label{sec:detailed_training_procedure}

During a training step, we sample a batch of $B$ samples $\left\{(V_{i_1}, y_{i_1}), ..., (V_{i_B}, y_{i_B})\right\}$ with replacement from our training dataset. For each sample, we sample a unit vector $\uv n$ from the uniform distribution over the unit sphere. We then extract $M_\mathrm{train}$ slices of shape $224 \times 224$ (${h_S = w_S = 224}$) from that volume with normal vector $\uv n$, offsets $s_m$ ranging from -1 to 1, and random vectors $\mathbf{u}_m, \mathbf{v}_m \in \uv n^\perp, m \in \left\{1, ..., M_\mathrm{train}\right\}$. $\mathbf u_m$ and $\mathbf v_m$ are chosen perpendicular to each other. Their lengths are randomly and independently chosen in the range ${[1 - 0.3, 1 + 0.3]}$, which amounts to applying random anisotropic scale augmentations. Furthermore, we introduce slice-wise random translation in the range $(-0.3, 0.3)$ and affine intensity augmentations where the values of the pixels are shifted according to the function $y = ax + b$ with $a, b \sim \mathcal{U}(-0.3, 0.3)$.
We concatenate the $B \times M_\mathrm{train}$ slices in a batch and perform a training step by associating to each slice the label of the volume from which it came.
In our experiments, we set $B=2$ and $M_\mathrm{train}=120$. Furthermore, we always sample one positive and one negative volume per batch, as we observe that this improves learning.
We use the binary cross-entropy loss and optimize the neural network with Adam \cite{kingma2014adam} with $\beta_1 = 0.9$, $\beta_2 = 0.999$, $\epsilon = 10^{-8}$. We use a cosine learning rate schedule \cite{loshchilov2016sgdr} starting at $5\cdot 10^{-5}$ and ending a $10^{-7}$, which updates the learning rate at each training step.
The number of training iterations for the Multiple Sclerosis, AutoPET, MosMed and Duke datasets are respectively set to 250,000, 187,500, 125,000 and 62,500. During training, we monitor the performance of the network on the validation set by obtaining volume-level predictions by max-pooling the slice level predictions. At the end of training, the validation balanced accuracies for the ResNet-10-T with frozen batch normalization where as follows: Multiple Sclerosis 0.93, AutoPET 0.88, MosMed 0.92, Duke 0.82.

\subsection{Evaluation details}

\label{sec:binarization_and_metrics}

\paragraph{Heatmap binarization}
In order to compute segmentation metrics, we first need to binarize the heatmaps that we obtain with the different methods.
We start by extracting the maximum value of each heatmap and find the dataset-dependent threshold $\tau$ that maximizes the weighted balanced accuracy:
$$wBA = \frac{5 \cdot  \mathrm{sensitivity} + \mathrm{specificity}}{6}$$
where the predicted label for a sample is considered to be positive if the maximum value of the heatmap is above the threshold $\tau$ and negative otherwise. We give more weight to sensitivity because for medical applications, sensitivity is often more important than specificity: it may be more problematic for a tumour to go undetected than to detect something that is not a tumour. We tried to use this threshold $\tau$ to directly binarize the heatmaps, but this resulted in undersegmentation.
We thus use a more advanced binarization procedure. First, we binarize the heatmaps with a threshold $\tau' < \tau$. Then, we extract each connected component, and reject the connected components whose maximum value in the heatmap is less than $\tau$. This way, we obtain a better segmentation without changing the predicted binary label of a given sample. We grid-search the value of $\tau'$ that maximizes the dice score between the predicted and ground truth segmentations, using 5 ground truth segmentation masks per dataset. For the Duke dataset, which only has one bounding box annotation per positive sample, we pick the value of $\tau'$ such that the intersection over union (IoU) of the ground truth bounding box and the best matching predicted bounding box is maximized.

We chose not to optimize $\tau$ and $\tau'$ on the training set because the models risk being overfitted. We instead perform 10-fold Monte-Carlo cross-validation. For each fold, we optimize $\tau$ and $\tau'$ on the first half of the shuffled validation dataset (for optimizing $\tau'$ we randomly select 5 samples with ground truth mask/bounding box in this first half) and compute the metrics on the second half. The results that we report are averaged over the 10 folds.

\paragraph{Metrics}
For the Multiple Sclerosis, AutoPET and MosMed datasets, ground truth segmentations are available, which allows to compute the dice score with respect to the predicted segmentation. We then compute the connected components of the predicted and ground truth segmentations. We consider that a connected component in the prediction is a true positive if there is a connected component in the ground truth that has an IoU with it greater than $1/8$ (for reference, an IoU of $1/8$ corresponds to two overlapping spheres with respective radiuses $r$ and $2r$). Otherwise, it is considered a false positive. Similarly, a connected component in the ground truth is considered a false negative if no connected component in the prediction has IoU with it greater than $1/8$.
On each sample with non-empty ground truth segmentation, we then compute the precision, recall and F1-score. These three metrics, in addition to the dice score, are averaged over the different samples in the fold.

For the Duke breast cancer dataset, only one bounding box annotation per patient is available. We start by computing the connected components of the predicted segmentation. We then obtain the bounding boxes of the different connected components and compare them to the unique ground truth bounding box. We apply the same definitions of true positives, false positive and false negatives as above, using IoU between bounding boxes instead of IoU between segmentation masks. Instead of the dice score, we report the maximum IoU between the ground truth bounding box and any predicted bounding box. As only at most one bounding box is available for each patient, despite there sometimes being multiple tumours, many predicted connected components that are considered as false positives may actually be true positives. Thus, precision and F1-score are less relevant for this dataset.

For all datasets, we also report the global balanced accuracy, as defined in the previous paragraph, but with equal weights for sensitivity and specificity.
We do not compute precision and F1-score on samples with empty ground truth segmentation. False positives in these samples thus have no effect on these metrics, but they are captured by the global balanced accuracy, which will be impacted if too many false positives are predicted for negative samples.

\subsection{Results for additional baselines}
\label{sec:rebuttal_additions}
In \Cref{tab:rebuttal_additions}, we additionally provide results for ScoreCAM \cite{score_cam} and GradCAM++ \cite{gradcam_pp}. We also provide results where the CAM methods are applied along all three spatial axes (i.e. axial, coronal and sagittal) and then averaged (denoted by a $\bigstar$). As in the rest of the paper, all CAM methods were evaluated once for each layer (1 to 4) and the results presented in \Cref{tab:rebuttal_additions} are the best out of the four layers.

\begin{table*}[htb]
    \centering
    \begin{tabular}{llccc}

   {\bf  Dataset } &{\bf   Method }& \bf{F1-score} & \bf{Dice/Max IoU} & {\bf Balanced accuracy} \\
    \toprule
    \multirow{11}{*}{\shortstack[l]{AutoPET}}
        & GradCAM & 0.06 & 0.11 & 0.63 \\
        & GradCAM $\bigstar$ & 0.05 & 0.15 & 0.65 \\
        & GradCAM++ & 0.22 & 0.21 & 0.61 \\
        & GradCAM++ $\bigstar$ & 0.18 & 0.18 & 0.57 \\
        & LayerCAM & 0.35 & 0.31 & 0.80 \\
        & LayerCAM $\bigstar$ & 0.35 & 0.37 & 0.74 \\
        & ScoreCAM & 0.23 & 0.23 & 0.63 \\
        & ScoreCAM $\bigstar$ & 0.20 & 0.23 & 0.58 \\
        & ToNNO (ours) & 0.39 & 0.32 & 0.74 \\
        & Averaged LayerCAM (ours) & 0.40 & \textbf{0.40} & \textbf{0.83} \\
        & Tomographic LayerCAM (ours) & \textbf{0.49} & 0.39 & 0.74 \\
    \midrule
    \multirow{11}{*}{\shortstack[l]{MosMed}}
        & GradCAM & 0.23 & 0.24 & 0.78 \\
        & GradCAM $\bigstar$ & 0.31 & 0.31 & 0.90 \\
        & GradCAM++ & 0.29 & 0.28 & 0.81 \\
        & GradCAM++ $\bigstar$ & 0.26 & 0.38 & 0.89 \\
        & LayerCAM & 0.39 & 0.35 & 0.90 \\
        & LayerCAM $\bigstar$ & 0.48 & 0.44 & 0.84 \\
        & ScoreCAM & 0.29 & 0.27 & 0.60 \\
        & ScoreCAM $\bigstar$ & 0.38 & 0.38 & 0.85 \\
        & ToNNO (ours) & 0.50 & 0.35 & 0.93 \\
        & Averaged LayerCAM (ours) & 0.53 & \textbf{0.48} & \textbf{0.95} \\
        & Tomographic LayerCAM (ours) & \textbf{0.55} & 0.41 & 0.89 \\
    \midrule
    \multirow{11}{*}{\shortstack[l]{Duke}}
        & GradCAM & 0.07 & 0.15 & 0.51 \\
        & GradCAM $\bigstar$ & 0.08 & 0.14 & 0.55 \\
        & GradCAM++ & 0.09 & 0.21 & 0.57 \\
        & GradCAM++ $\bigstar$ & 0.34 & 0.30 & \textbf{0.80} \\
        & LayerCAM & 0.24 & 0.29 & 0.58 \\
        & LayerCAM $\bigstar$ & 0.44 & 0.38 & 0.74 \\
        & ScoreCAM & 0.18 & 0.27 & 0.53 \\
        & ScoreCAM $\bigstar$ & 0.23 & 0.20 & 0.60 \\
        & ToNNO (ours) & \textbf{0.51} & 0.37 & 0.79 \\
        & Averaged LayerCAM (ours) & 0.47 & \textbf{0.42} & 0.75 \\
        & Tomographic LayerCAM (ours) & \textbf{0.51} & \textbf{0.42} & 0.79 \\
    \midrule
    \end{tabular}
    \caption{In this table, we provide results that were requested by the reviewers. Unfortunately, because of the premature termination of a contract with the company providing the private Multiple Sclerosis dataset, we had to delete all data and were unable to run new experiments on this dataset. $\bigstar$ means averaging along the three directions (axial, coronal, sagittal).}
    \label{tab:rebuttal_additions}
    \vspace{-1.9em}
\end{table*}

\clearpage

\onecolumn

\subsection{Multiple Sclerosis dataset results}

\vfill

\begin{table}[htb]
    \centering
    \begin{tabular}{lccccccc}
    \toprule
    Method & Precision & Recall & F1-score & Dice & Balanced accuracy \\
    \toprule
    GradCAM (layer 1) & 0.04 & 0.27 & 0.06 & 0.04 & 0.57 \\
    GradCAM (layer 2) & 0.15 & 0.38 & 0.18 & 0.12 & 0.64 \\
    GradCAM (layer 3) & 0.03 & 0.02 & 0.02 & 0.06 & 0.86 \\
    GradCAM (layer 4) & 0.00 & 0.00 & 0.00 & 0.02 & 0.91 \\
    \midrule 
    LayerCAM (layer 1) & 0.76 & 0.78 & 0.73 & 0.41 & 0.89 \\
    LayerCAM (layer 2) & 0.51 & 0.50 & 0.48 & 0.24 & 0.91 \\
    LayerCAM (layer 3) & 0.02 & 0.01 & 0.02 & 0.07 & \textbf{0.94} \\
    LayerCAM (layer 4) & 0.00 & 0.00 & 0.00 & 0.02 & 0.91 \\
    \midrule 
    ToNNO & 0.82 & 0.88 & 0.82 & 0.53 & 0.90 \\
    \midrule 
    Averaged LayerCAM (layer 1) & 0.86 & 0.87 & \textbf{0.84} & 0.52 & 0.91 \\
    Averaged LayerCAM (layer 2) & 0.66 & 0.70 & 0.65 & 0.30 & 0.90 \\
    Averaged LayerCAM (layer 3) & 0.11 & 0.12 & 0.11 & 0.09 & 0.90 \\
    Averaged LayerCAM (layer 4) & 0.15 & 0.14 & 0.14 & 0.07 & 0.91 \\
    \midrule 
    Tomographic LayerCAM (layer 1) & 0.77 & 0.86 & 0.78 & 0.55 & 0.88 \\
    Tomographic LayerCAM (layer 2) & \textbf{0.87} & 0.87 & \textbf{0.84} & \textbf{0.58} & \textbf{0.94} \\
    Tomographic LayerCAM (layer 3) & 0.83 & \textbf{0.90} & \textbf{0.84} & 0.57 & 0.93 \\
    Tomographic LayerCAM (layer 4) & 0.81 & 0.87 & 0.81 & 0.53 & 0.91 \\
    \bottomrule
    \end{tabular}
    \caption{Results for the Multiple Sclerosis dataset.}
    \label{tab:gvf_results}
\end{table}

\vfill

\begin{figure}[htb]
    \centering
    \includegraphics[width=0.965\textwidth, trim={0 0cm 0 0cm}, clip]{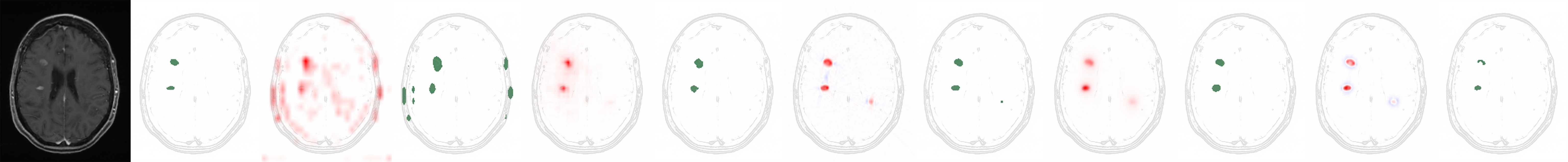} \\
    \includegraphics[width=0.965\textwidth, trim={0 0cm 0 0cm}, clip]{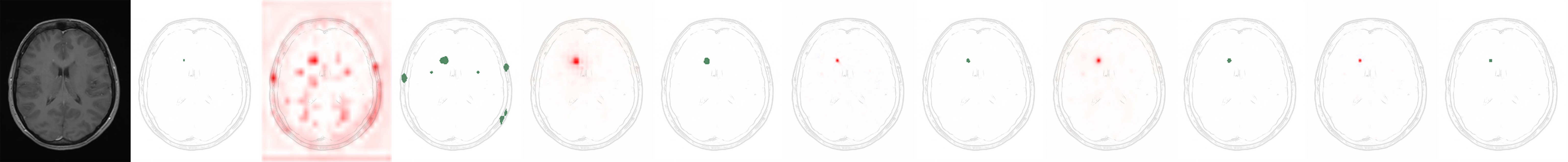} \\
    \includegraphics[width=0.965\textwidth, trim={0 0cm 0 0cm}, clip]{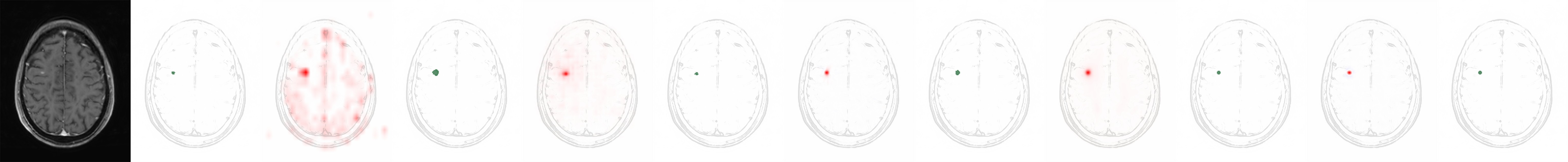} \\
    \includegraphics[width=0.965\textwidth, trim={0 0cm 0 0cm}, clip]{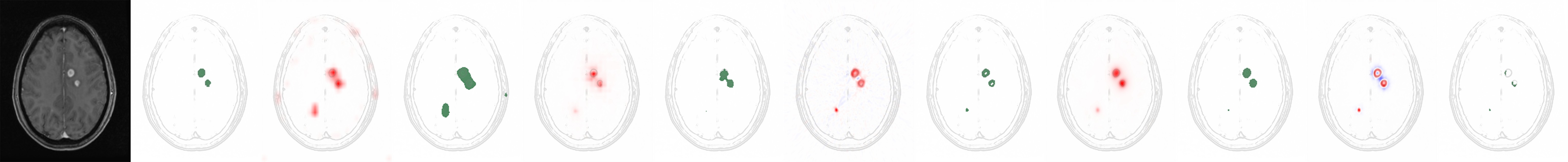} \\
    \includegraphics[width=0.965\textwidth, trim={0 0cm 0 0cm}, clip]{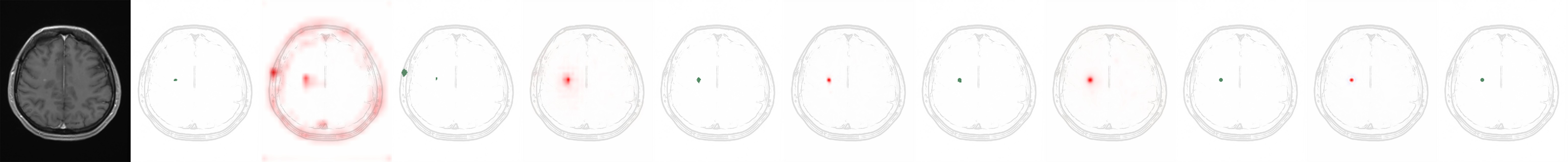} \\

    \captionsetup[subfigure]{labelformat=empty}
    \captionsetup[subfigure]{justification=centering}
    \captionsetup[subfigure]{font=figurefont}
    
    \begin{minipage}[t]{0.075\textwidth}\vspace{-10pt}
    \subcaption{Input}
    \end{minipage}
    \begin{minipage}[t]{0.075\textwidth}\vspace{-10pt}
    \subcaption{Ground truth}
    \end{minipage}
    \begin{minipage}[t]{0.075\textwidth}\vspace{-10pt}
    \subcaption{GradCAM}
    \end{minipage}
    \begin{minipage}[t]{0.075\textwidth}\vspace{-10pt}
    \subcaption{GradCAM\\binarized}
    \end{minipage}
    \begin{minipage}[t]{0.075\textwidth}\vspace{-10pt}
    \subcaption{LayerCAM}
    \end{minipage}
    \begin{minipage}[t]{0.075\textwidth}\vspace{-10pt}
    \subcaption{LayerCAM\\binarized}
    \end{minipage}
    \begin{minipage}[t]{0.075\textwidth}\vspace{-10pt}
    \subcaption{ToNNO}
    \end{minipage}
    \begin{minipage}[t]{0.075\textwidth}\vspace{-10pt}
    \subcaption{ToNNO binarized}
    \end{minipage}
    \begin{minipage}[t]{0.075\textwidth}\vspace{-10pt}
    \subcaption{Averaged LayerCAM}
    \end{minipage}
    \begin{minipage}[t]{0.075\textwidth}\vspace{-10pt}
    \subcaption{Averaged LayerCAM binarized}
    \end{minipage}
    \begin{minipage}[t]{0.075\textwidth}\vspace{-10pt}
    \subcaption{Tomographic LayerCAM}
    \end{minipage}
    \begin{minipage}[t]{0.075\textwidth}\vspace{-10pt}
    \subcaption{Tomographic LayerCAM binarized}
    \end{minipage}
    
    \caption{\label{fig:gvf_examples} Examples for the Multiple Sclerosis dataset}
\end{figure}

\vfill

\begin{figure}[htb]
    \centering

    \captionsetup[subfigure]{labelformat=empty}
    \captionsetup[subfigure]{justification=centering}
    \captionsetup[subfigure]{font=figurefont}
    
    \begin{minipage}[c]{0.05\textwidth}
        \subcaption{Layer 1}
    \end{minipage}
    \begin{minipage}[c]{0.945\textwidth}
        \includegraphics[width=\textwidth, trim={0 0cm 0 0cm}, clip]{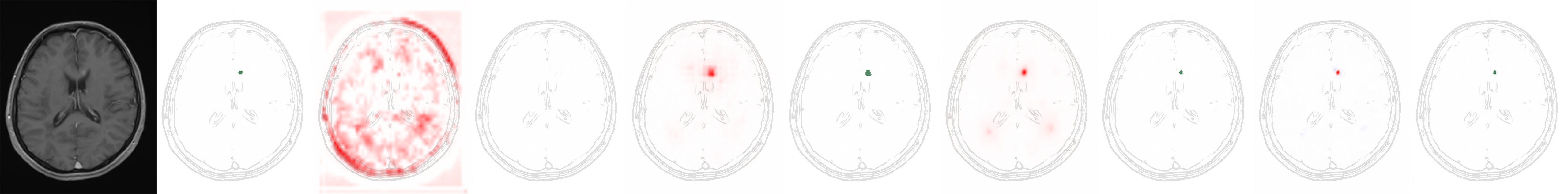}
    \end{minipage}

    \begin{minipage}[c]{0.05\textwidth}
        \subcaption{Layer 2}
    \end{minipage}
    \begin{minipage}[c]{0.945\textwidth}
        \includegraphics[width=\textwidth, trim={0 0cm 0 0cm}, clip]{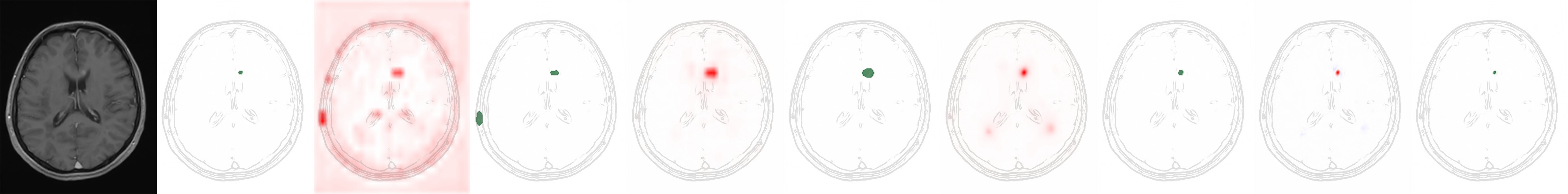}
    \end{minipage}

    \begin{minipage}[c]{0.05\textwidth}
        \subcaption{Layer 3}
    \end{minipage}
    \begin{minipage}[c]{0.945\textwidth}
        \includegraphics[width=\textwidth, trim={0 0cm 0 0cm}, clip]{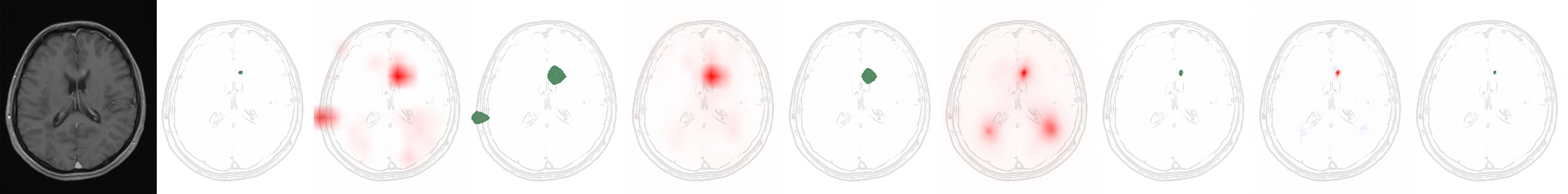}
    \end{minipage}

    \begin{minipage}[c]{0.05\textwidth}
        \subcaption{Layer 4}
    \end{minipage}
    \begin{minipage}[c]{0.945\textwidth}
        \includegraphics[width=\textwidth, trim={0 0cm 0 0cm}, clip]{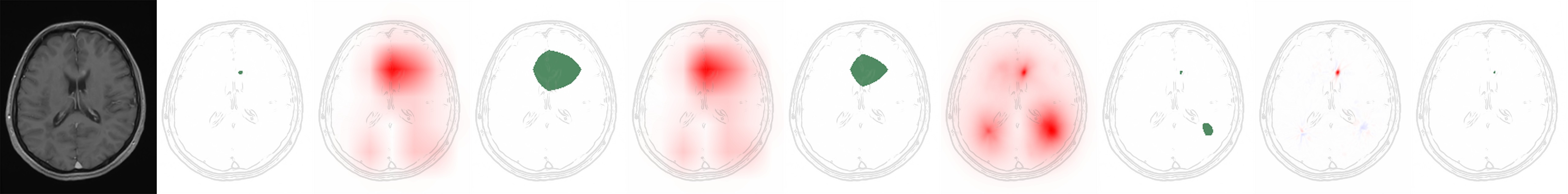}
    \end{minipage}

    \begin{minipage}[t]{0.045\textwidth}\vspace{-0pt}
        \subcaption{}
    \end{minipage}
    \begin{minipage}[t]{0.089\textwidth}\vspace{-0pt}
        \subcaption{Input}
    \end{minipage}
    \begin{minipage}[t]{0.089\textwidth}\vspace{-0pt}
        \subcaption{Ground truth}
    \end{minipage}
    \begin{minipage}[t]{0.089\textwidth}\vspace{-0pt}
        \subcaption{GradCAM}
    \end{minipage}
    \begin{minipage}[t]{0.089\textwidth}\vspace{-0pt}
        \subcaption{GradCAM\\binarized}
    \end{minipage}
    \begin{minipage}[t]{0.089\textwidth}\vspace{-0pt}
        \subcaption{LayerCAM}
    \end{minipage}
    \begin{minipage}[t]{0.089\textwidth}\vspace{-0pt}
        \subcaption{LayerCAM\\binarized}
    \end{minipage}
    \begin{minipage}[t]{0.089\textwidth}\vspace{-0pt}
        \subcaption{Averaged LayerCAM}
    \end{minipage}
    \begin{minipage}[t]{0.089\textwidth}\vspace{-0pt}
        \subcaption{Averaged LayerCAM binarized}
    \end{minipage}
    \begin{minipage}[t]{0.089\textwidth}\vspace{-0pt}
        \subcaption{Tomographic LayerCAM}
    \end{minipage}
    \begin{minipage}[t]{0.089\textwidth}\vspace{-0pt}
        \subcaption{Tomographic LayerCAM binarized}
    \end{minipage}
    
    \caption{\label{fig:gvf_layers1to4} Output of GradCAM, LayerCAM, Averaged LayerCAM and Tomographic LayerCAM for layers 1 to 4 for one sample of the Multiple Sclerosis dataset.}
\end{figure}

\vfill

\clearpage

\subsection{AutoPET dataset results}

\vfill

\begin{table*}[htb]
    \centering
    \begin{tabular}{lccccccc}
    \toprule
    Method & Precision & Recall & F1-score & Dice & Balanced accuracy \\
    \toprule

    GradCAM (layer 1) & 0.00 & 0.00 & 0.00 & 0.01 & 0.51 \\
    GradCAM (layer 2) & 0.04 & 0.04 & 0.02 & 0.03 & 0.55 \\
    GradCAM (layer 3) & 0.10 & 0.10 & 0.06 & 0.11 & 0.63 \\
    GradCAM (layer 4) & 0.04 & 0.04 & 0.03 & 0.10 & 0.87 \\
    \midrule 
    LayerCAM (layer 1) & 0.43 & 0.30 & 0.29 & 0.28 & 0.69 \\
    LayerCAM (layer 2) & 0.59 & 0.29 & 0.35 & 0.31 & 0.80 \\
    LayerCAM (layer 3) & 0.26 & 0.14 & 0.16 & 0.20 & 0.84 \\
    LayerCAM (layer 4) & 0.05 & 0.04 & 0.04 & 0.11 & \textbf{0.89} \\
    \midrule 
    ToNNO & 0.52 & 0.41 & 0.39 & 0.32 & 0.74 \\
    \midrule 
    Averaged LayerCAM (layer 1) & 0.68 & 0.34 & 0.38 & 0.37 & 0.82 \\
    Averaged LayerCAM (layer 2) & 0.65 & 0.33 & 0.40 & \textbf{0.40} & 0.83 \\
    Averaged LayerCAM (layer 3) & 0.44 & 0.21 & 0.25 & 0.24 & 0.81 \\
    Averaged LayerCAM (layer 4) & 0.19 & 0.07 & 0.09 & 0.08 & 0.78 \\
    \midrule 
    Tomographic LayerCAM (layer 1) & 0.68 & \textbf{0.47} & \textbf{0.49} & 0.39 & 0.74 \\
    Tomographic LayerCAM (layer 2) & \textbf{0.71} & 0.41 & 0.46 & 0.38 & 0.79 \\
    Tomographic LayerCAM (layer 3) & 0.66 & 0.39 & 0.42 & 0.34 & 0.81 \\
    Tomographic LayerCAM (layer 4) & 0.61 & 0.46 & 0.46 & 0.32 & 0.77 \\
    \bottomrule
    \end{tabular}
    \caption{Results for the AutoPET dataset.}
    \label{tab:autopet_results}
\end{table*}

\vfill

\begin{figure*}[htb]
    \centering
    \includegraphics[width=0.965\textwidth, trim={0 4cm 0 1cm}, clip]{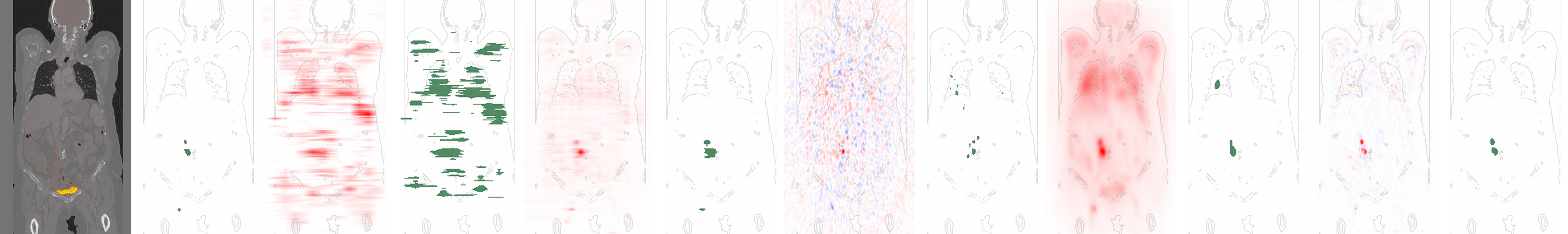} \\
    \includegraphics[width=0.965\textwidth, trim={0 4cm 0 1cm}, clip]{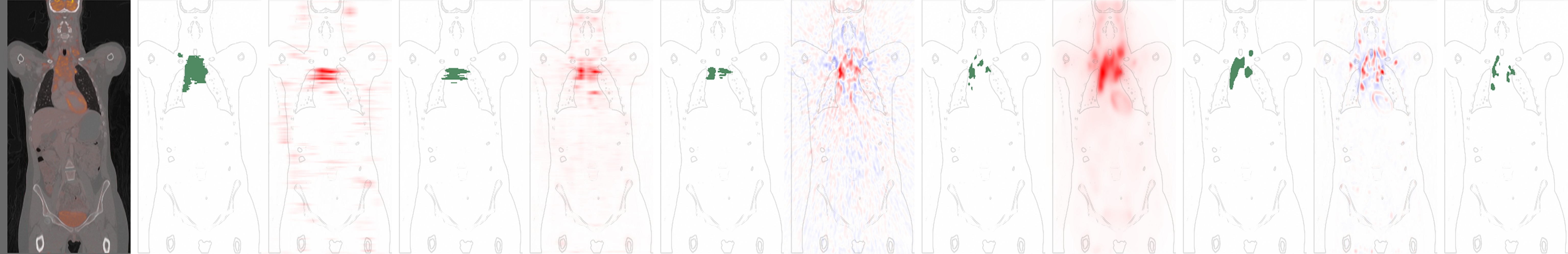} \\
    \includegraphics[width=0.965\textwidth, trim={0 4cm 0 1cm}, clip]{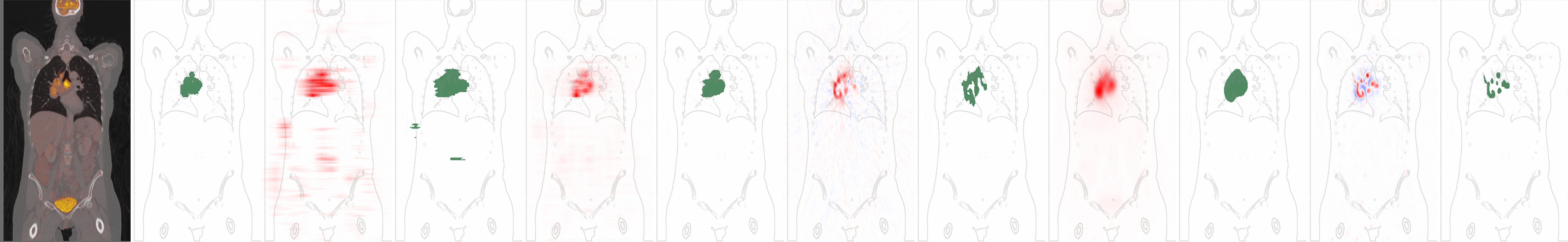} \\
    \includegraphics[width=0.965\textwidth, trim={0 4cm 0 1cm}, clip]{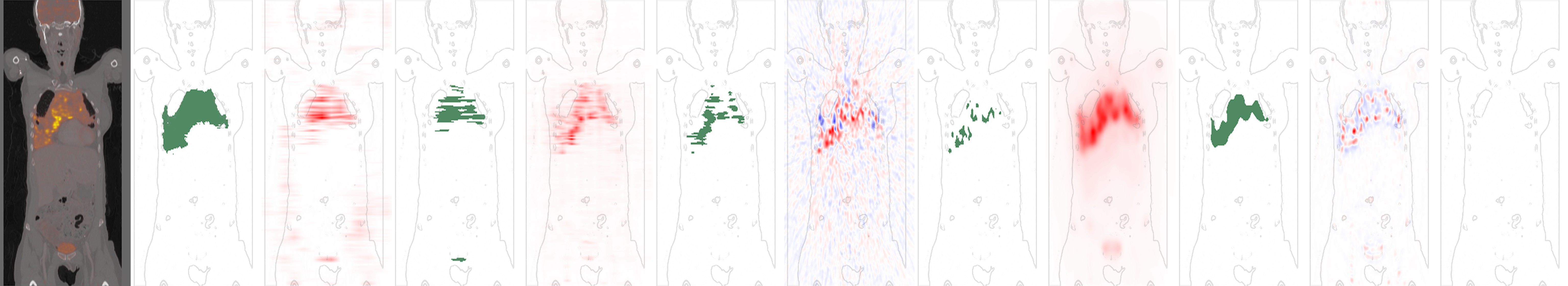} \\
    
    \captionsetup[subfigure]{labelformat=empty}
    \captionsetup[subfigure]{justification=centering}
    \captionsetup[subfigure]{font=figurefont}
    
    \begin{minipage}[t]{0.075\textwidth}\vspace{-10pt}
    \subcaption{Input}
    \end{minipage}
    \begin{minipage}[t]{0.075\textwidth}\vspace{-10pt}
    \subcaption{Ground truth}
    \end{minipage}
    \begin{minipage}[t]{0.075\textwidth}\vspace{-10pt}
    \subcaption{GradCAM}
    \end{minipage}
    \begin{minipage}[t]{0.075\textwidth}\vspace{-10pt}
    \subcaption{GradCAM\\binarized}
    \end{minipage}
    \begin{minipage}[t]{0.075\textwidth}\vspace{-10pt}
    \subcaption{LayerCAM}
    \end{minipage}
    \begin{minipage}[t]{0.075\textwidth}\vspace{-10pt}
    \subcaption{LayerCAM\\binarized}
    \end{minipage}
    \begin{minipage}[t]{0.075\textwidth}\vspace{-10pt}
    \subcaption{ToNNO}
    \end{minipage}
    \begin{minipage}[t]{0.075\textwidth}\vspace{-10pt}
    \subcaption{ToNNO binarized}
    \end{minipage}
    \begin{minipage}[t]{0.075\textwidth}\vspace{-10pt}
    \subcaption{Averaged LayerCAM}
    \end{minipage}
    \begin{minipage}[t]{0.075\textwidth}\vspace{-10pt}
    \subcaption{Averaged LayerCAM binarized}
    \end{minipage}
    \begin{minipage}[t]{0.075\textwidth}\vspace{-10pt}
    \subcaption{Tomographic LayerCAM}
    \end{minipage}
    \begin{minipage}[t]{0.075\textwidth}\vspace{-10pt}
    \subcaption{Tomographic LayerCAM binarized}
    \end{minipage}
    
    \caption{\label{fig:autopet_examples} Examples for the AutoPET dataset}
\end{figure*}

\vfill

\begin{figure}[htb]
    \centering

    \captionsetup[subfigure]{labelformat=empty}
    \captionsetup[subfigure]{justification=centering}
    \captionsetup[subfigure]{font=figurefont}
    
    \begin{minipage}[c]{0.05\textwidth}
        \subcaption{Layer 1}
    \end{minipage}
    \begin{minipage}[c]{0.945\textwidth}
        \includegraphics[width=\textwidth, trim={0 0cm 0 0cm}, clip]{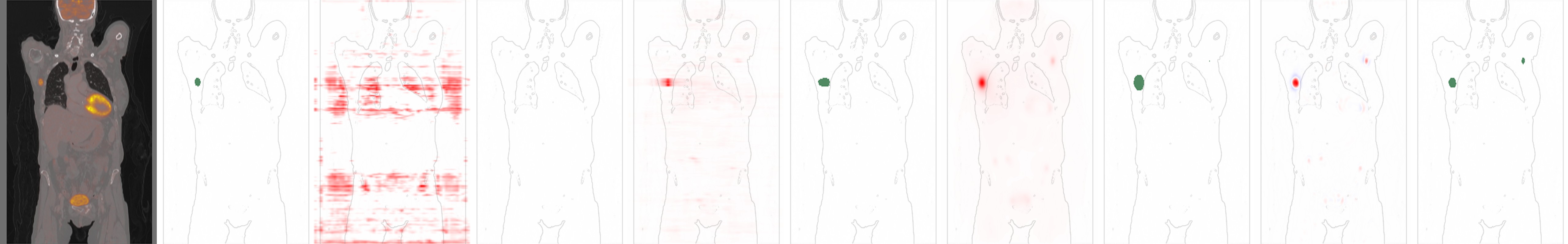}
    \end{minipage}

    \begin{minipage}[c]{0.05\textwidth}
        \subcaption{Layer 2}
    \end{minipage}
    \begin{minipage}[c]{0.945\textwidth}
        \includegraphics[width=\textwidth, trim={0 0cm 0 0cm}, clip]{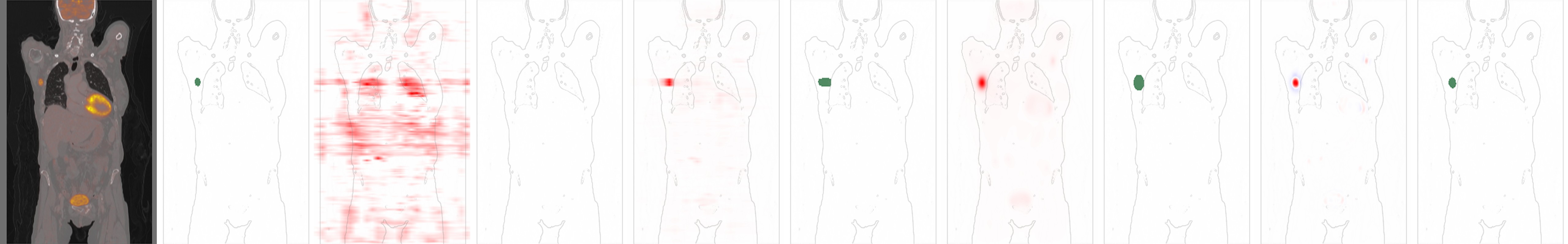}
    \end{minipage}

    \begin{minipage}[c]{0.05\textwidth}
        \subcaption{Layer 3}
    \end{minipage}
    \begin{minipage}[c]{0.945\textwidth}
        \includegraphics[width=\textwidth, trim={0 0cm 0 0cm}, clip]{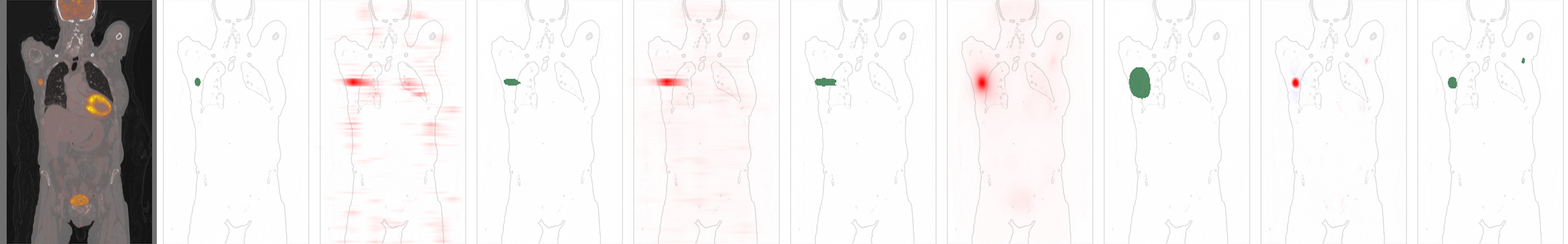}
    \end{minipage}

    \begin{minipage}[c]{0.05\textwidth}
        \subcaption{Layer 4}
    \end{minipage}
    \begin{minipage}[c]{0.945\textwidth}
        \includegraphics[width=\textwidth, trim={0 0cm 0 0cm}, clip]{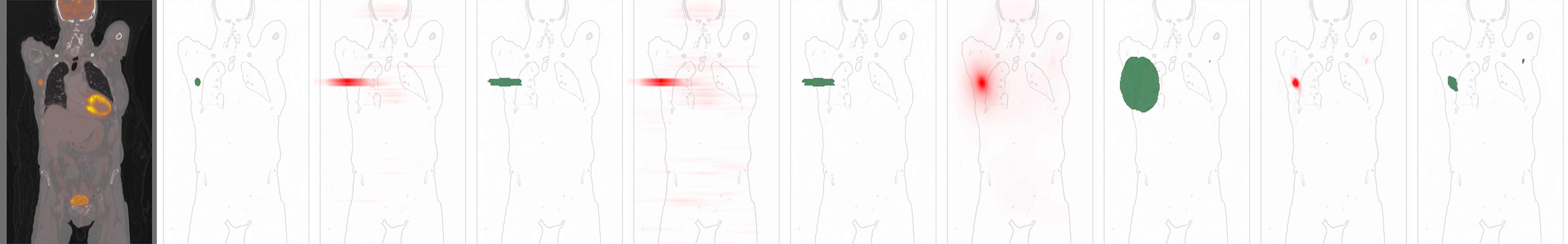}
    \end{minipage}

    \begin{minipage}[t]{0.045\textwidth}\vspace{-0pt}
        \subcaption{}
    \end{minipage}
    \begin{minipage}[t]{0.089\textwidth}\vspace{-0pt}
        \subcaption{Input}
    \end{minipage}
    \begin{minipage}[t]{0.089\textwidth}\vspace{-0pt}
        \subcaption{Ground truth}
    \end{minipage}
    \begin{minipage}[t]{0.089\textwidth}\vspace{-0pt}
        \subcaption{GradCAM}
    \end{minipage}
    \begin{minipage}[t]{0.089\textwidth}\vspace{-0pt}
        \subcaption{GradCAM\\binarized}
    \end{minipage}
    \begin{minipage}[t]{0.089\textwidth}\vspace{-0pt}
        \subcaption{LayerCAM}
    \end{minipage}
    \begin{minipage}[t]{0.089\textwidth}\vspace{-0pt}
        \subcaption{LayerCAM\\binarized}
    \end{minipage}
    \begin{minipage}[t]{0.089\textwidth}\vspace{-0pt}
        \subcaption{Averaged LayerCAM}
    \end{minipage}
    \begin{minipage}[t]{0.089\textwidth}\vspace{-0pt}
        \subcaption{Averaged LayerCAM binarized}
    \end{minipage}
    \begin{minipage}[t]{0.089\textwidth}\vspace{-0pt}
        \subcaption{Tomographic LayerCAM}
    \end{minipage}
    \begin{minipage}[t]{0.089\textwidth}\vspace{-0pt}
        \subcaption{Tomographic LayerCAM binarized}
    \end{minipage}
    
    \caption{\label{fig:autopet_layers1to4} Output of GradCAM, LayerCAM, Averaged LayerCAM and Tomographic LayerCAM for layers 1 to 4 for one sample of the AutoPET dataset.}
\end{figure}

\vfill

\clearpage

\subsection{MosMed dataset results}

\vfill

\begin{table*}[htb]
    \centering
    \begin{tabular}{lccccccc}
    \toprule
    Method & Precision & Recall & F1-score & Dice & Balanced accuracy \\
    \toprule

    GradCAM (layer 1) & 0.00 & 0.00 & 0.00 & 0.01 & 0.57 \\
    GradCAM (layer 2) & 0.02 & 0.02 & 0.01 & 0.02 & 0.64 \\
    GradCAM (layer 3) & 0.35 & 0.25 & 0.23 & 0.24 & 0.78 \\
    GradCAM (layer 4) & 0.22 & 0.12 & 0.14 & 0.17 & 0.87 \\
    \midrule 
    LayerCAM (layer 1) & 0.52 & 0.25 & 0.29 & 0.28 & 0.85 \\
    LayerCAM (layer 2) & 0.68 & 0.32 & 0.39 & 0.35 & 0.90 \\
    LayerCAM (layer 3) & 0.45 & 0.28 & 0.30 & 0.28 & 0.78 \\
    LayerCAM (layer 4) & 0.23 & 0.13 & 0.15 & 0.17 & 0.87 \\
    \midrule 
    ToNNO & 0.69 & 0.45 & 0.50 & 0.35 & 0.93 \\
    \midrule 
    Averaged LayerCAM (layer 1) & 0.70 & 0.44 & 0.49 & 0.41 & 0.90 \\
    Averaged LayerCAM (layer 2) & \textbf{0.72} & 0.49 & 0.53 & \textbf{0.48} & \textbf{0.95} \\
    Averaged LayerCAM (layer 3) & 0.68 & 0.37 & 0.43 & 0.37 & \textbf{0.95} \\
    Averaged LayerCAM (layer 4) & 0.32 & 0.16 & 0.19 & 0.17 & 0.94 \\
    \midrule 
    Tomographic LayerCAM (layer 1) & 0.57 & 0.41 & 0.43 & 0.29 & 0.76 \\
    Tomographic LayerCAM (layer 2) & 0.68 & \textbf{0.52} & \textbf{0.55} & 0.41 & 0.87 \\
    Tomographic LayerCAM (layer 3) & \textbf{0.72} & \textbf{0.52} & \textbf{0.55} & 0.41 & 0.89 \\
    Tomographic LayerCAM (layer 4) & 0.71 & 0.46 & 0.52 & 0.38 & 0.92 \\
    \bottomrule
    \end{tabular}
    \caption{Results for the MosMed dataset.}
    \label{tab:mosmed_results}
\end{table*}

\vfill

\begin{figure*}[htb]
    \centering
    \includegraphics[width=0.955\textwidth, trim={0 0cm 0 0cm}, clip]{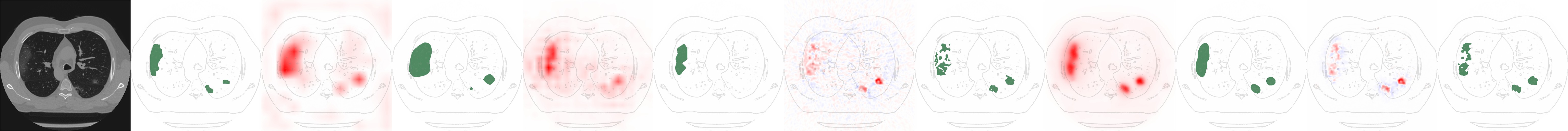} \\
    \includegraphics[width=0.955\textwidth, trim={0 0cm 0 0cm}, clip]{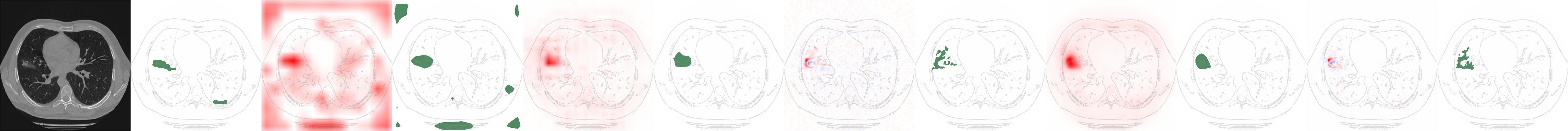} \\
    \includegraphics[width=0.955\textwidth, trim={0 0cm 0 0cm}, clip]{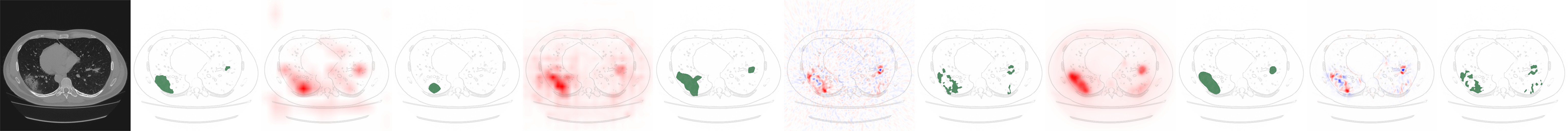} \\
    \includegraphics[width=0.955\textwidth, trim={0 0cm 0 0cm}, clip]{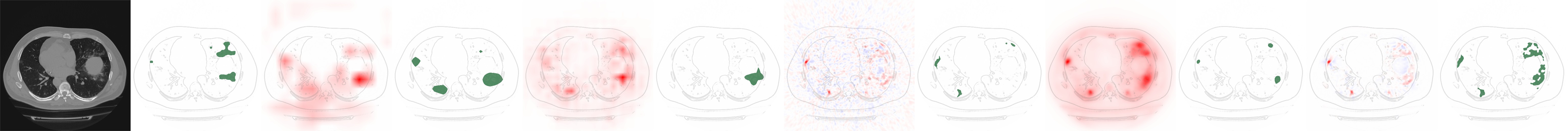} \\
    \includegraphics[width=0.955\textwidth, trim={0 0cm 0 0cm}, clip]{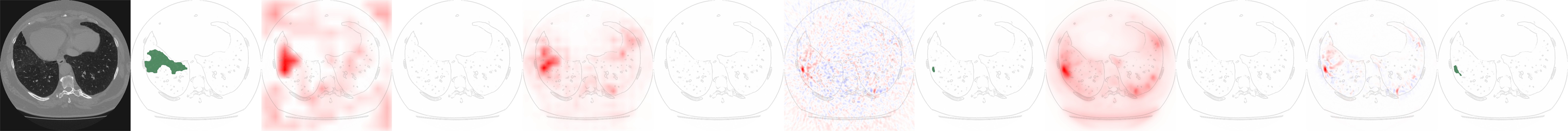} \\
    \includegraphics[width=0.955\textwidth, trim={0 0cm 0 0cm}, clip]{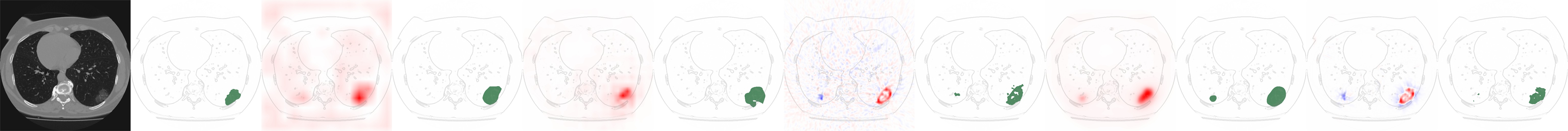} \\

    \captionsetup[subfigure]{labelformat=empty}
    \captionsetup[subfigure]{justification=centering}
    \captionsetup[subfigure]{font=figurefont}
    
    \begin{minipage}[t]{0.075\textwidth}\vspace{-10pt}
    \subcaption{Input}
    \end{minipage}
    \begin{minipage}[t]{0.075\textwidth}\vspace{-10pt}
    \subcaption{Ground truth}
    \end{minipage}
    \begin{minipage}[t]{0.075\textwidth}\vspace{-10pt}
    \subcaption{GradCAM}
    \end{minipage}
    \begin{minipage}[t]{0.075\textwidth}\vspace{-10pt}
    \subcaption{GradCAM\\binarized}
    \end{minipage}
    \begin{minipage}[t]{0.075\textwidth}\vspace{-10pt}
    \subcaption{LayerCAM}
    \end{minipage}
    \begin{minipage}[t]{0.075\textwidth}\vspace{-10pt}
    \subcaption{LayerCAM\\binarized}
    \end{minipage}
    \begin{minipage}[t]{0.075\textwidth}\vspace{-10pt}
    \subcaption{ToNNO}
    \end{minipage}
    \begin{minipage}[t]{0.075\textwidth}\vspace{-10pt}
    \subcaption{ToNNO binarized}
    \end{minipage}
    \begin{minipage}[t]{0.075\textwidth}\vspace{-10pt}
    \subcaption{Averaged LayerCAM}
    \end{minipage}
    \begin{minipage}[t]{0.075\textwidth}\vspace{-10pt}
    \subcaption{Averaged LayerCAM binarized}
    \end{minipage}
    \begin{minipage}[t]{0.075\textwidth}\vspace{-10pt}
    \subcaption{Tomographic LayerCAM}
    \end{minipage}
    \begin{minipage}[t]{0.075\textwidth}\vspace{-10pt}
    \subcaption{Tomographic LayerCAM binarized}
    \end{minipage}
    
    \caption{\label{fig:mosmed_examples} Examples for the MosMed COVID-19 thoracic CT dataset}
\end{figure*}

\vfill

\begin{figure}[htb]
    \centering

    \captionsetup[subfigure]{labelformat=empty}
    \captionsetup[subfigure]{justification=centering}
    \captionsetup[subfigure]{font=figurefont}
    
    \begin{minipage}[c]{0.05\textwidth}
        \subcaption{Layer 1}
    \end{minipage}
    \begin{minipage}[c]{0.945\textwidth}
        \includegraphics[width=\textwidth, trim={0 0cm 0 0cm}, clip]{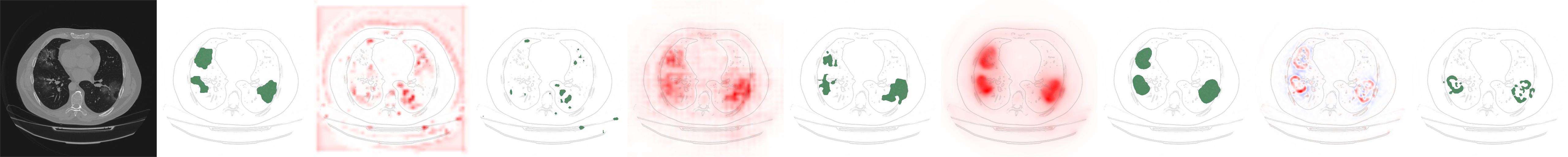}
    \end{minipage}

    \begin{minipage}[c]{0.05\textwidth}
        \subcaption{Layer 2}
    \end{minipage}
    \begin{minipage}[c]{0.945\textwidth}
        \includegraphics[width=\textwidth, trim={0 0cm 0 0cm}, clip]{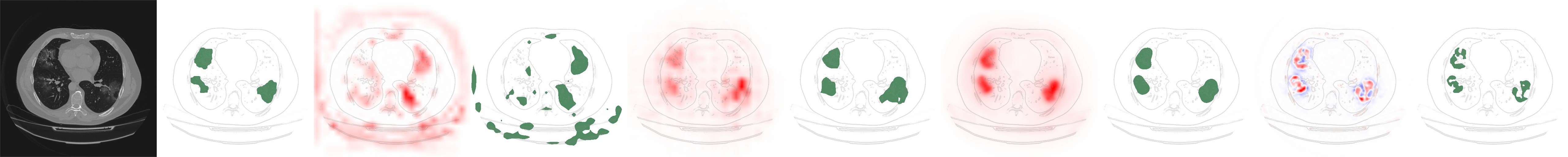}
    \end{minipage}

    \begin{minipage}[c]{0.05\textwidth}
        \subcaption{Layer 3}
    \end{minipage}
    \begin{minipage}[c]{0.945\textwidth}
        \includegraphics[width=\textwidth, trim={0 0cm 0 0cm}, clip]{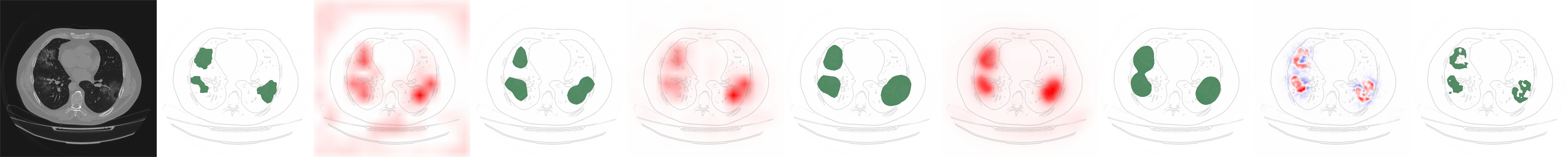}
    \end{minipage}

    \begin{minipage}[c]{0.05\textwidth}
        \subcaption{Layer 4}
    \end{minipage}
    \begin{minipage}[c]{0.945\textwidth}
        \includegraphics[width=\textwidth, trim={0 0cm 0 0cm}, clip]{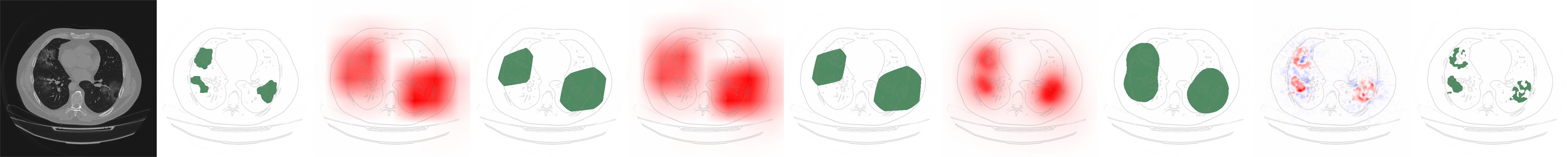}
    \end{minipage}

    \begin{minipage}[t]{0.045\textwidth}\vspace{-0pt}
        \subcaption{}
    \end{minipage}
    \begin{minipage}[t]{0.089\textwidth}\vspace{-0pt}
        \subcaption{Input}
    \end{minipage}
    \begin{minipage}[t]{0.089\textwidth}\vspace{-0pt}
        \subcaption{Ground truth}
    \end{minipage}
    \begin{minipage}[t]{0.089\textwidth}\vspace{-0pt}
        \subcaption{GradCAM}
    \end{minipage}
    \begin{minipage}[t]{0.089\textwidth}\vspace{-0pt}
        \subcaption{GradCAM\\binarized}
    \end{minipage}
    \begin{minipage}[t]{0.089\textwidth}\vspace{-0pt}
        \subcaption{LayerCAM}
    \end{minipage}
    \begin{minipage}[t]{0.089\textwidth}\vspace{-0pt}
        \subcaption{LayerCAM\\binarized}
    \end{minipage}
    \begin{minipage}[t]{0.089\textwidth}\vspace{-0pt}
        \subcaption{Averaged LayerCAM}
    \end{minipage}
    \begin{minipage}[t]{0.089\textwidth}\vspace{-0pt}
        \subcaption{Averaged LayerCAM binarized}
    \end{minipage}
    \begin{minipage}[t]{0.089\textwidth}\vspace{-0pt}
        \subcaption{Tomographic LayerCAM}
    \end{minipage}
    \begin{minipage}[t]{0.089\textwidth}\vspace{-0pt}
        \subcaption{Tomographic LayerCAM binarized}
    \end{minipage}
    
    \caption{\label{fig:mosmed_layers1to4} Output of GradCAM, LayerCAM, Averaged LayerCAM and Tomographic LayerCAM for layers 1 to 4 for one sample of the MosMed dataset.}
\end{figure}
\vfill

\clearpage

\subsection{Duke dataset results}

\vfill

\begin{table*}[htb]
    \centering
    \begin{tabular}{lccccccc}
    \toprule
    Method & Precision & Recall & F1-score & MaxIoU & Balanced accuracy \\
    \toprule

    GradCAM (layer 1) & \textcolor{black}{0.00 }& 0.07 & \textcolor{black}{0.01 }& 0.03 & 0.54 \\
    GradCAM (layer 2) & \textcolor{black}{0.01 }& 0.11 & \textcolor{black}{0.01 }& 0.04 & 0.52 \\
    GradCAM (layer 3) & \textcolor{black}{0.04 }& 0.53 & \textcolor{black}{0.07 }& 0.15 & 0.51 \\
    GradCAM (layer 4) & \textcolor{black}{0.09 }& 0.22 & \textcolor{black}{0.11 }& 0.08 & 0.65 \\
    \midrule 
    LayerCAM (layer 1) & \textcolor{black}{0.10 }& 0.76 & \textcolor{black}{0.16 }& 0.29 & 0.53 \\
    LayerCAM (layer 2) & \textcolor{black}{0.17 }& 0.81 & \textcolor{black}{0.24 }& 0.29 & 0.58 \\
    LayerCAM (layer 3) & \textcolor{black}{0.16 }& 0.58 & \textcolor{black}{0.23 }& 0.17 & 0.59 \\
    LayerCAM (layer 4) & \textcolor{black}{0.11 }& 0.25 & \textcolor{black}{0.13 }& 0.08 & 0.65 \\
    \midrule 
    ToNNO & \textcolor{black}{\textbf{0.43} }& 0.77 & \textcolor{black}{\textbf{0.51} }& 0.37 & 0.79 \\
    \midrule 
    Averaged LayerCAM (layer 1) & \textcolor{black}{0.39 }& 0.83 & \textcolor{black}{0.49 }& 0.40 & 0.79 \\
    Averaged LayerCAM (layer 2) & \textcolor{black}{0.37 }& \textbf{0.84} & \textcolor{black}{0.47 }& \textbf{0.42} & 0.75 \\
    Averaged LayerCAM (layer 3) & \textcolor{black}{0.34 }& 0.59 & \textcolor{black}{0.40 }& 0.21 & 0.74 \\
    Averaged LayerCAM (layer 4) & \textcolor{black}{0.16 }& 0.22 & \textcolor{black}{0.18 }& 0.09 & 0.71 \\
    \midrule 
    Tomographic LayerCAM (layer 1) & \textcolor{black}{0.15 }& 0.82 & \textcolor{black}{0.23 }& 0.39 & 0.55 \\
    Tomographic LayerCAM (layer 2) & \textcolor{black}{0.30 }& 0.83 & \textcolor{black}{0.40 }& 0.40 & 0.68 \\
    Tomographic LayerCAM (layer 3) & \textcolor{black}{0.41 }& 0.81 & \textcolor{black}{\textbf{0.51} }& \textbf{0.42} & 0.79 \\
    Tomographic LayerCAM (layer 4) & \textcolor{black}{0.42 }& 0.80 & \textcolor{black}{\textbf{0.51} }& 0.39 & \textbf{0.80} \\

    \bottomrule
    \end{tabular}
    \caption{Results for the Duke dataset. Precision and F1-score are not as relevant as only at most one tumour per patient is reported in the ground truth data.}
    \label{tab:duke_results}
\end{table*}

\vfill

\begin{figure*}[htb]
    \centering
    \includegraphics[width=0.955\textwidth, trim={0 5cm 0 0cm}, clip]{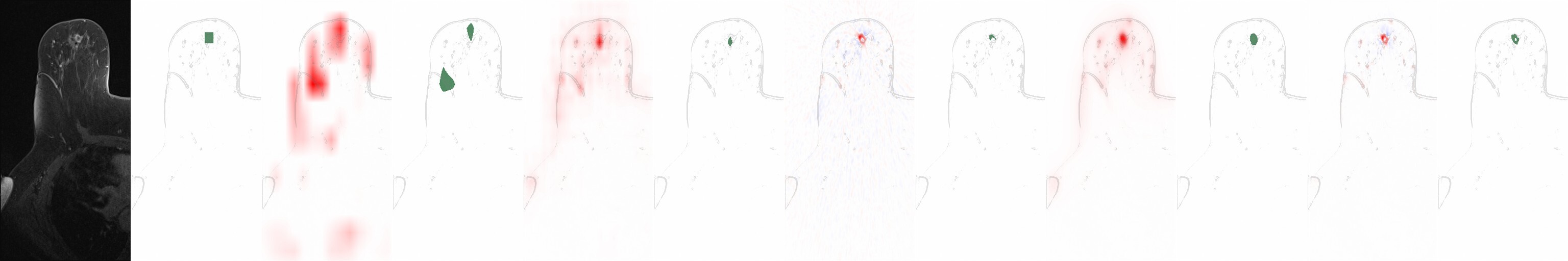} \\
    \includegraphics[width=0.955\textwidth, trim={0 5cm 0 0cm}, clip]{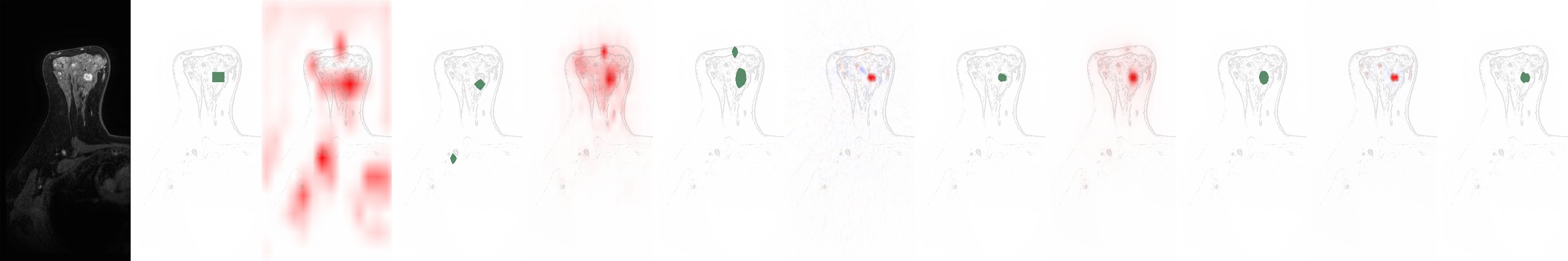} \\
    \includegraphics[width=0.955\textwidth, trim={0 5cm 0 0cm}, clip]{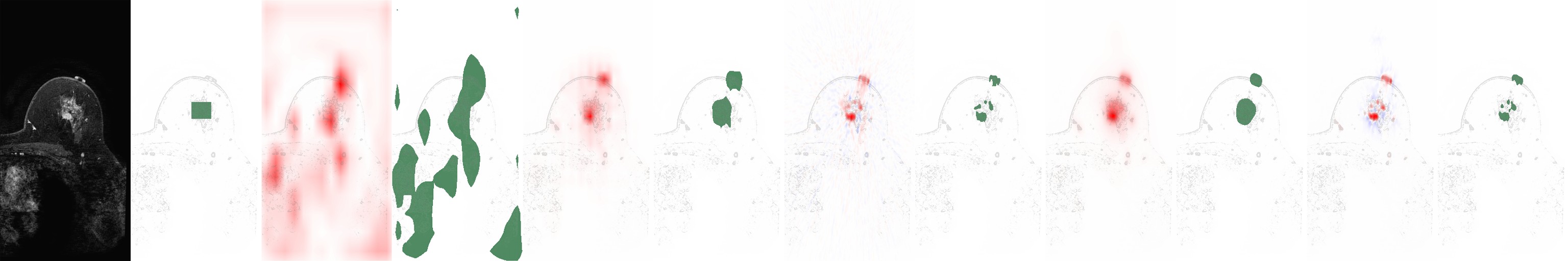} \\
    \includegraphics[width=0.955\textwidth, trim={0 5cm 0 0cm}, clip]{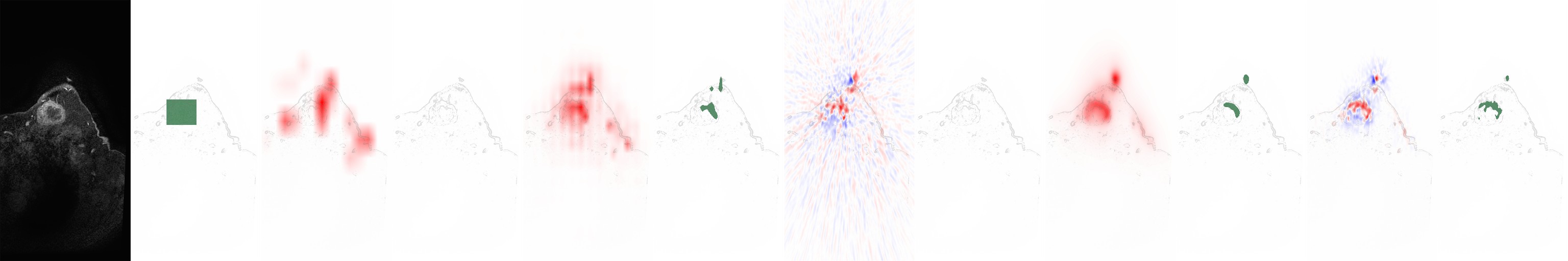} \\

    \captionsetup[subfigure]{labelformat=empty}
    \captionsetup[subfigure]{justification=centering}
    \captionsetup[subfigure]{font=figurefont}
    
    \begin{minipage}[t]{0.075\textwidth}\vspace{-10pt}
    \subcaption{Input}
    \end{minipage}
    \begin{minipage}[t]{0.075\textwidth}\vspace{-10pt}
    \subcaption{Ground truth}
    \end{minipage}
    \begin{minipage}[t]{0.075\textwidth}\vspace{-10pt}
    \subcaption{GradCAM}
    \end{minipage}
    \begin{minipage}[t]{0.075\textwidth}\vspace{-10pt}
    \subcaption{GradCAM\\binarized}
    \end{minipage}
    \begin{minipage}[t]{0.075\textwidth}\vspace{-10pt}
    \subcaption{LayerCAM}
    \end{minipage}
    \begin{minipage}[t]{0.075\textwidth}\vspace{-10pt}
    \subcaption{LayerCAM\\binarized}
    \end{minipage}
    \begin{minipage}[t]{0.075\textwidth}\vspace{-10pt}
    \subcaption{ToNNO}
    \end{minipage}
    \begin{minipage}[t]{0.075\textwidth}\vspace{-10pt}
    \subcaption{ToNNO binarized}
    \end{minipage}
    \begin{minipage}[t]{0.075\textwidth}\vspace{-10pt}
    \subcaption{Averaged LayerCAM}
    \end{minipage}
    \begin{minipage}[t]{0.075\textwidth}\vspace{-10pt}
    \subcaption{Averaged LayerCAM binarized}
    \end{minipage}
    \begin{minipage}[t]{0.075\textwidth}\vspace{-10pt}
    \subcaption{Tomographic LayerCAM}
    \end{minipage}
    \begin{minipage}[t]{0.075\textwidth}\vspace{-10pt}
    \subcaption{Tomographic LayerCAM binarized}
    \end{minipage}
    
    \caption{\label{fig:duke_examples} Examples for the Duke breast cancer MRI dataset}
\end{figure*}

\vfill

\begin{figure}[htb]
    \centering

    \captionsetup[subfigure]{labelformat=empty}
    \captionsetup[subfigure]{justification=centering}
    \captionsetup[subfigure]{font=figurefont}
    
    \begin{minipage}[c]{0.05\textwidth}
        \subcaption{Layer 1}
    \end{minipage}
    \begin{minipage}[c]{0.945\textwidth}
        \includegraphics[width=\textwidth, trim={0 0cm 0 0cm}, clip]{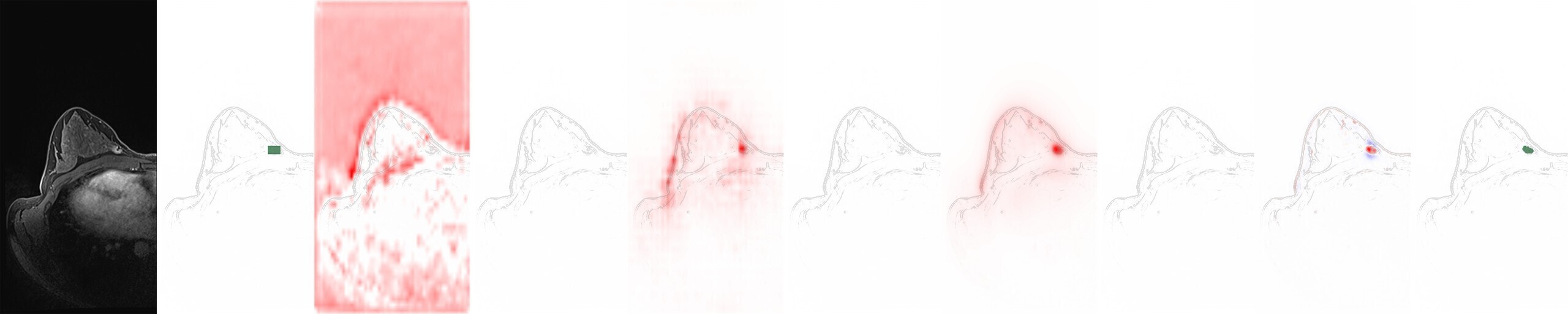}
    \end{minipage}

    \begin{minipage}[c]{0.05\textwidth}
        \subcaption{Layer 2}
    \end{minipage}
    \begin{minipage}[c]{0.945\textwidth}
        \includegraphics[width=\textwidth, trim={0 0cm 0 0cm}, clip]{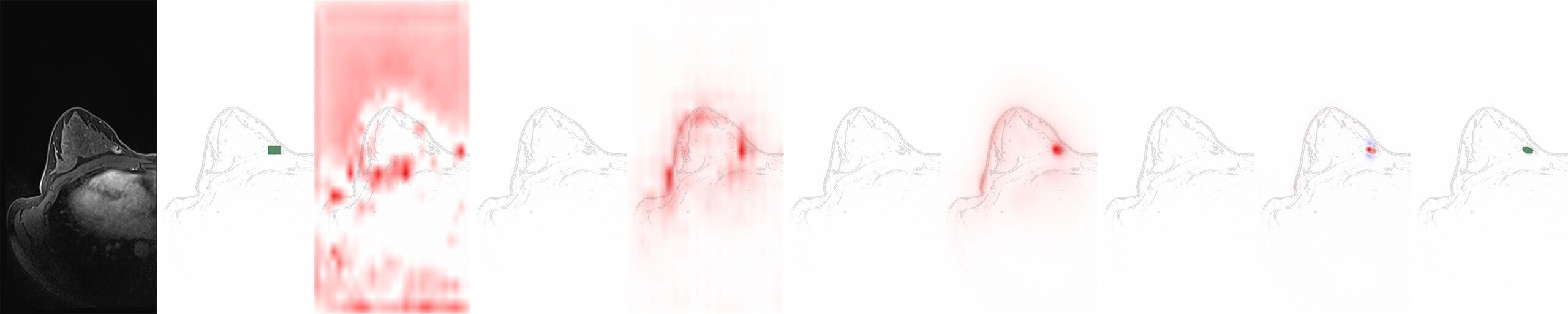}
    \end{minipage}

    \begin{minipage}[c]{0.05\textwidth}
        \subcaption{Layer 3}
    \end{minipage}
    \begin{minipage}[c]{0.945\textwidth}
        \includegraphics[width=\textwidth, trim={0 0cm 0 0cm}, clip]{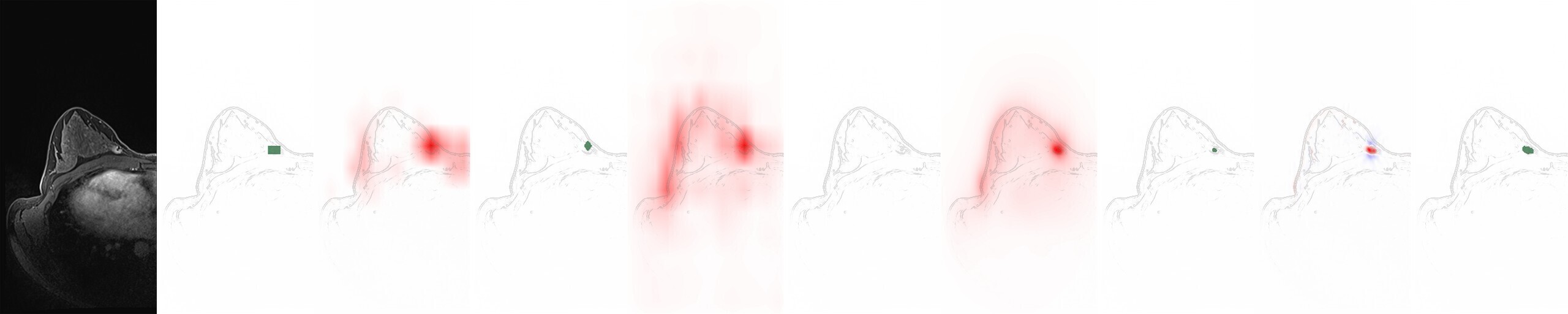}
    \end{minipage}

    \begin{minipage}[c]{0.05\textwidth}
        \subcaption{Layer 4}
    \end{minipage}
    \begin{minipage}[c]{0.945\textwidth}
        \includegraphics[width=\textwidth, trim={0 0cm 0 0cm}, clip]{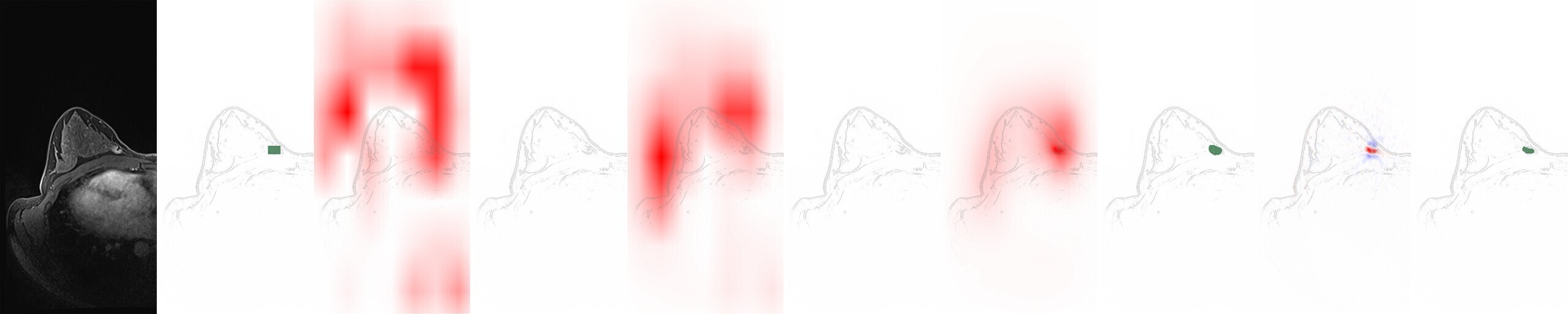}
    \end{minipage}

    \begin{minipage}[t]{0.045\textwidth}\vspace{-0pt}
        \subcaption{}
    \end{minipage}
    \begin{minipage}[t]{0.089\textwidth}\vspace{-0pt}
        \subcaption{Input}
    \end{minipage}
    \begin{minipage}[t]{0.089\textwidth}\vspace{-0pt}
        \subcaption{Ground truth}
    \end{minipage}
    \begin{minipage}[t]{0.089\textwidth}\vspace{-0pt}
        \subcaption{GradCAM}
    \end{minipage}
    \begin{minipage}[t]{0.089\textwidth}\vspace{-0pt}
        \subcaption{GradCAM\\binarized}
    \end{minipage}
    \begin{minipage}[t]{0.089\textwidth}\vspace{-0pt}
        \subcaption{LayerCAM}
    \end{minipage}
    \begin{minipage}[t]{0.089\textwidth}\vspace{-0pt}
        \subcaption{LayerCAM\\binarized}
    \end{minipage}
    \begin{minipage}[t]{0.089\textwidth}\vspace{-0pt}
        \subcaption{Averaged LayerCAM}
    \end{minipage}
    \begin{minipage}[t]{0.089\textwidth}\vspace{-0pt}
        \subcaption{Averaged LayerCAM binarized}
    \end{minipage}
    \begin{minipage}[t]{0.089\textwidth}\vspace{-0pt}
        \subcaption{Tomographic LayerCAM}
    \end{minipage}
    \begin{minipage}[t]{0.089\textwidth}\vspace{-0pt}
        \subcaption{Tomographic LayerCAM binarized}
    \end{minipage}
    
    \caption{\label{fig:duke_layers1to4} Output of GradCAM, LayerCAM, Averaged LayerCAM and Tomographic LayerCAM for layers 1 to 4 for one sample of the Duke dataset.}
\end{figure}

\vfill